\documentclass[aps,prc,twocolumn,superscriptaddress,showpacs,floatfix]{revtex4-1}
\usepackage{graphicx}

\usepackage{bm}
\usepackage{setspace}
\usepackage{lineno}
\usepackage{multirow}
\usepackage{hyperref}
\usepackage[usenames]{color}

\newcommand{\meanpt}{$\left\langle p_T\right\rangle$} 
\newcommand{\pt}{$p_{\rm T}$}

\begin{document}

\title[Hadron Suppression At BRAHMS]{Rapidity and centrality dependence of particle production for identified hadrons in Cu+Cu collisions at $\sqrt{s_{NN}} = 200$ GeV}

\newcommand{\bnl}{Brookhaven National Laboratory, Upton, NY}
\newcommand{\krakow}{~Smoluchowski Inst. of Physics, Jagiellonian University, Krakow, Poland}
 \newcommand{\newyork}{~New York University, New York, NY}
 \newcommand{\nbi}{Niels Bohr Institute, Blegdamsvej 17, University of Copenhagen, Copenhagen 2100, Denmark}
 \newcommand{\texas}{Texas A$\&$M University, College Station, Texas}
 \newcommand{\bergen}{~University of Bergen, Department of Physics and Technology, Bergen, Norway}
 \newcommand{\bucharest}{~University of Bucharest, Bucharest, Romania}
 \newcommand{\kansas}{~University of Kansas, Lawrence, Kansas}
 \newcommand{\oslo}{~niversity of Oslo, Department of Physics, Oslo, Norway}
\newcommand{\spacescience}{~Institute for Space Sciences, Bucharest, Romania}
 \newcommand{\iphc}{Institute Pluridisciplinaire Hubert Curien CRNS-IN2P3 et Universit{\'e} de Strasbourg, Strasbourg, France}

\affiliation{\bnl}
\affiliation{\iphc}
\affiliation{Institue of Space Science, Bucharest-Magurele, Romania}
\affiliation{M. Smoluchowski Inst. of Physics, Jagiellonian University, Krakow, Poland}
\affiliation{New York University, New York}
\affiliation{Niels Bohr Institute, University of Copenhagen, Copenhagen, Denmark}
\affiliation{Texas A\&M University, College Station, Texas}
\affiliation{University of Bergen, Department of Physics and Technology, Bergen, Norway}
\affiliation{University of Bucharest, Romania}
\affiliation{The University Of Kansas, Lawrence, Kansas}
\affiliation{University of Oslo, Department of Physics, Oslo, Norway}

\author{I.~C.~Arsene}\affiliation{University of Oslo, Department of Physics, Oslo, Norway}
\author{I.~G.~Bearden}\affiliation{Niels Bohr Institute, University of Copenhagen, Copenhagen, Denmark}
\author{D.~Beavis}\affiliation{\bnl}
\author{S.~Bekele}\altaffiliation[Present address: ]{Department of Polymer Science, The University of Akron, Akron Ohio}
\affiliation{The University Of Kansas, Lawrence, Kansas}
\author{C.~Besliu}\affiliation{University of Bucharest, Romania}
\author{B.~Budick}\affiliation{New York University, New York}
\author{H.~B{\o}ggild}\affiliation{Niels Bohr Institute, University of Copenhagen, Copenhagen, Denmark}
\author{C.~Chasman}\affiliation{\bnl}
\author{C.~H.~Christensen}\affiliation{Niels Bohr Institute, University of Copenhagen, Copenhagen, Denmark}
\author{P.~Christiansen}\altaffiliation[Present Address: ]{Div. of Experimental
High-Energy Physics, Lund University, Lund, Sweden}\affiliation{Niels Bohr Institute, University of Copenhagen, Copenhagen, Denmark}
\author{H.~H.~Dalsgaard}\affiliation{Niels Bohr Institute, University of Copenhagen, Copenhagen, Denmark}
\author{R.~Debbe}\affiliation{\bnl}
\author{J.~J.~Gaardh{\o}je}\affiliation{Niels Bohr Institute, University of Copenhagen, Copenhagen, Denmark}
\author{K.~Hagel}\affiliation{Texas A\&M University, College Station, Texas}
\author{H.~Ito}\affiliation{\bnl}
\author{A.~Jipa}\affiliation{University of Bucharest, Romania}
\author{E.~B.~Johnson}\altaffiliation[Present Address]{Radiation Monitoring Devices,  Cambridge, MA, USA}\affiliation{The University Of Kansas, Lawrence, Kansas}
\author{C.~E.~J{\o}rgensen}\altaffiliation[Present address: ]{Ris{\o} National Laboratory, Denmark}\affiliation{Niels Bohr Institute, University of Copenhagen, Copenhagen, Denmark}
\author{R.~Karabowicz}\affiliation{M. Smoluchowski Inst. of Physics, Jagiellonian University, Krakow, Poland}
\author{N.~Katrynska}\affiliation{M. Smoluchowski Inst. of Physics, Jagiellonian University, Krakow, Poland}
\author{E.~J.~Kim}\altaffiliation[Present address: ]{Division of Science Education, Chonbuk National University, Jeonju, 561-756, Korea}\affiliation{The University Of Kansas, Lawrence, Kansas}
\author{T.~M.~Larsen}\affiliation{Niels Bohr Institute, University of Copenhagen, Copenhagen, Denmark}
\author{J.~H.~Lee}\affiliation{\bnl}
\author{G.~L{\o}vh{\o}iden}\affiliation{University of Oslo, Department of Physics, Oslo, Norway}
\author{Z.~Majka}\affiliation{M. Smoluchowski Inst. of Physics, Jagiellonian University, Krakow, Poland}
\author{M.J.~Murray}\affiliation{The University Of Kansas, Lawrence, Kansas}
\author{J.~Natowitz}\affiliation{Texas A\&M University, College Station, Texas}
\author{B.S.~Nielsen}\affiliation{Niels Bohr Institute, University of Copenhagen, Copenhagen, Denmark}
\author{C.~Nygaard}\affiliation{Niels Bohr Institute, University of Copenhagen, Copenhagen, Denmark}
\author{D.~Pal}\affiliation{The University Of Kansas, Lawrence, Kansas}
\author{A.~Qviller}\affiliation{University of Oslo, Department of Physics, Oslo, Norway}
\author{F.~Rami}\affiliation{\iphc}
\author{C.~Ristea}\affiliation{Institue of Space Science, Bucharest-Magurele, Romania}
\author{O.~Ristea}\affiliation{University of Bucharest, Romania}
\author{D.~R{\"o}hrich}\affiliation{University of Bergen, Department of Physics and Technology, Bergen, Norway}
\author{S.~J.~Sanders}\affiliation{The University Of Kansas, Lawrence, Kansas}
\author{P.~Stazel}\affiliation{M. Smoluchowski Inst. of Physics, Jagiellonian University, Krakow, Poland}
\author{T.~S.~Tveter}\affiliation{University of Oslo, Department of Physics, Oslo, Norway}
\author{F.~Videb{\ae}k}\altaffiliation[]{Spokesperson}\affiliation{\bnl}
\author{R.~Wada}\altaffiliation[Present address]{Institute of Modern Physics, Chinese Academy of Sciences, Lanzhou, China}\affiliation{Texas A\&M University, College Station, Texas}
\author{H.~Yang}\affiliation{University of Bergen, Department of Physics and Technology, Bergen, Norway}
\author{Z.~Yin}\altaffiliation[Present address: ]{Huazhong Normal University,Wuhan, China}\affiliation{University of Bergen, Department of Physics and Technology, Bergen, Norway}
\author{I.~S.~Zgura}\affiliation{Institue of Space Science, Bucharest-Magurele, Romania}

\date{\today}

\setstretch{1.5}

\begin{abstract}
The BRAHMS collaboration has measured transverse momentum spectra of  pions, kaons, protons and antiprotons at rapidities 0 and 3 for Cu+Cu collisions at $\sqrt{s_{NN}} = 200$ GeV. 
As the collisions become more central the collective radial flow increases while  the  temperature of kinetic freeze-out decreases. 
The temperature is lower and the radial flow weaker at forward rapidity. 
Pion and kaon yields with transverse momenta between 1.5 and 2.5\,GeV/$c$ are suppressed for central collisions relative to scaled $p+p$ collisions. 
This suppression, which increases as the collisions become more central is consistent with  jet quenching models
and is also present  with comparable magnitude at forward rapidity. 
At such rapidities initial state effects may also be present and persistence of the meson suppression to high rapidity may reflect a combination of jet quenching and nuclear shadowing. 
The ratio of protons to mesons increases as the collisions become more central and is largest at forward rapidities.

\end{abstract}

\pacs{25.75.Gz}

\maketitle

\section{Introduction}
\newcounter{mycounter1}[section]

Collisions of ions at the Relativistic Heavy Ion Collider (RHIC) with masses as heavy as Au and
center of mass energies of 200 GeV per nucleon to produce
extended systems that have been characterized as being partonic, strongly coupled and 
exhibiting hydrodynamic flow behavior with a viscosity per degree of freedom  near 
the theoretical lower limit \cite{Kovtun:2004de}.
This  medium is known as the  strongly coupled  Quark Gluon Plasma or sQGP \cite{Arsene:2004fa,Adcox:2004mh,Back:2004je,Adams:2005dq}. 

The matter created in these heavy ion collisions exists for a very short period of time
as it expands and cools down with the subsequent hadronization of all partons, some of which
are eventually detected by the experiments as jets or leading hadrons. 
The medium can be explored by comparing spectra of hard probes from heavy-ion collisions (where the partons have to traverse an extended medium) 
to those of a smaller 
system, such as $p+p$ collisions, at the same energy  per nucleon. 
Jet and leading hadron measurements are believed to probe the early stages of the dense medium while soft 
hadronic observables deliver information on the 
initial state and hydrodynamic evolution of the system. 

The systematic study of such observables as a function of the number of participants in the collisions N$_{\rm part}$,   has been very important in understanding the matter created in Au+Au collisions at $\sqrt{s_{NN}}=200$\,GeV. 
However for peripheral Au+Au collisions with $N_{\rm part} < 60 $ the uncertainties on $N_{\rm part}$ are of the order of 20\% \cite{Adler:2003cb,Adare:2013esx} leaving room for different scenarios for the dependence
of particle production on the system size. 

In order to extend the medium size dependence of physical observables down to 
small systems such as d+Au  and $p+p$, the Cu+Cu  system, with $A_{Cu} =63$, was selected since it 
provides a good overlap with peripheral Au+Au collisions in terms of the number of participants. 
The relative  uncertainty in the fractional cross-section of Cu+Cu collisions is
 smaller compared to that  in Au+Au collisions for the same number of participants.
Assuming a uniform mass distribution, the overlap region in central Cu+Cu collisions is spherical while that in Au+Au collisions for the same number
 of participants has an almond shape, making it possible to explore geometry effects on
 the experimental observables. 
 The core/corona model of K. Werner \cite{Werner:2007bf} and Beccattini and 
Manninen \cite{Becattini:2008ya} provides a mechanism for testing these effects since the ratio of core
 to corona depends upon the shape of the overlap region. 
 
Most available data of identified hadrons are from near mid-rapidity. 
The BRAHMS data offers a unique opportunity to study hadron production at both mid and forward rapidity and compare properties to further enhance our knowledge of the matter formed and different chemical conditions. 
In this paper, we present transverse-momentum spectra, yields,  mean transverse momenta $\langle p_{T} \rangle$, nuclear modification
factors ($R_{AA}$) and ratios for identified charged hadrons ($\pi^{\pm}, K^{\pm},p,\bar{p}$) 
obtained with the BRAHMS spectrometers in Cu+Cu collisions
at $\sqrt{s_{NN}}$ = 200 GeV.  
The measurements were done at two rapidities $y = 0$ and $y = 3$ as a function of collision centrality. 
Blast wave fits to the transverse momentum spectra are used to extract the mean transverse velocity and kinetic temperature at the kinetic freeze-out point.
The results are compared to those obtained in $p+p$ and Au+Au
collisions at the same energy, rapidity and centrality (number of participants) where available.

\section{The BRAHMS Experiment}

The BRAHMS Experiment consists of two small acceptance magnetic spectrometers, 
the Mid-Rapidity Spectrometer (MRS) and the Forward Spectrometer (FS), for 
tracking, momentum determination, and particle identification  together with a system of global detectors made up of 
Beam-Beam Counters (BBC), Zero Degree Calorimeters (ZDCs) and a Multiplicity Array (MA)
 \cite{Adamczyk:2003sq,Adler:2000bd}. The global detectors are used for triggering, centrality determination, and
 separating nuclear from electromagnetic events.
The MRS uses two time projection chambers (TPCs), TPM1 and TPM2, with a magnet between 
them and time of flight (TOF) walls for particle identification (PID). 
The Forward Spectrometer (FS) has two TPCs (T1 and T2) and three Drift Chambers (DCs) 
with magnets located between the  detectors. In the FS, PID is achieved by using a TOF 
wall behind T2 and a second TOF wall and a Ring Imaging Cherenkov (RICH) detector
both placed after the third DC. 
 The TPCs and DCs each provide several three dimensional space points which together
 with the momentum information provided by the deflection in magnets allow for particle tracking.
The MRS is capable of rotating between $90^{\circ}$ and $30^{\circ}$ with respect to the beam
pipe covering the rapidity interval from $y \sim 0$ to $y \sim 1.6$. The FS rotates
between $15^{\circ}$ and $2^{\circ}$ and covers the rapidity interval from $y \sim 2.2$ to 
$y \sim 4.0$. 
For the data presented in this paper, the MRS was set at $90^{\circ}$ and the FS was set at $4^{\circ}$. These settings correspond to $y=0$ and  $y \sim 3$, respectively.

The primary collision vertex position is determined  to an accuracy of $\sim$1 cm based on the 
relative time-of-flight of fast ($\beta \approx 1$) particles hitting the  beam-beam counter arrays (BBC).  
The BBCs consist of Cherenkov detectors mounted on photomultiplier tubes and are located 220~cm 
from the nominal vertex position on either side of the interaction region. 
The BBCs also provide the start time for the time of flight (TOF) measurements.

\subsection{Event Selection}

The centrality of the collisions is characterized by using  a multiplicity array (MA), which consists of an inner layer of Si strip detectors 
and an outer layer of scintillator tiles each arranged as hexagonal barrels coaxial with the beam pipe. 
By measuring the energy loss of charged particles that traverse the two arrays,
the strip detectors and the tiles provide two semi-independent
 measurements from which the pseudo-rapidity dependence of the
 charged particle density can be deduced.
A realistic GEANT3 simulation of the detector response is used in this
determination to map energy  
deposits 
to the corresponding number of primary 
particles \cite{Brun:1994aa}. 
Reaction centrality is based on the distribution of 
charged particle multiplicities within the nominal pseudo-rapidity range covered by 
the MA,  $|\eta| < 2.2$. 

For a given event the centrality was taken to be defined as the fraction of observed events with a greater integral of charged particle multiplicity than that event. 
With this definition, $0\%$ centrality--correspond to collisions with the greatest  overlap of the two nuclei. 
Events generated by HIJING were passed through a GEANT3 simulation of the experiment and used to estimate the number of peripheral  events missed because they do not leave sufficient energy in the MA for detection.
The procedure applied for determining centrality and the associated numbers of participants, $\langle N_{part} \rangle$, and binary nucleon-nucleon collisions, $\langle N_{coll} \rangle$, in the Cu+Cu system is the same as described in detail for the Au+Au analysis \cite{Bearden:2001qq}. 
The values extracted from this procedure are displayed in Table~\ref{tab:npart}.

\begin{table}[ht]
\begin{tabular}{|c|c|c|}
\hline
Cent. & $\langle N_{part} \rangle$ & $\langle N_{coll} \rangle$  \\ \hline
~0-10\% & $97 \pm 0.8$ & 166 $\pm$ 2 \\ 
10-30\% & 61 $\pm$ 2.6 & ~85 $\pm$ 5 \\
30-50\% & 29 $\pm$ 4.3 & ~30 $\pm$ 6 \\
50-70\% & 12 $\pm$ 3.2 & ~~9.6 $\pm$ 3.2 \\ \hline
\end{tabular}
\caption{$\langle N_{part} \rangle$ and $\langle N_{coll} \rangle$ for the centrality ranges used for Cu+Cu in this paper. Note the errors are correlated between different centrality values.}
\label{tab:npart}
\end{table}

For this analysis, the events were divided into four centrality 
classes ($0-10\%,10-30\%,30-50\% ~\rm{and} ~ 50-70\%$). 
Events within $\pm 25$~cm of the nominal vertex were selected. 
Since the spectrometer acceptance depends upon the location of the vertex for a given event,  
spectral analysis is carried out in vertex bins of 5~cm and the results 
are statistically averaged to obtain the final spectra.

\subsection{Track Selection}

Straight line track segments are determined by  tracking detectors, which  are outside the magnetic field regions. 
These 
track segment are joined inside the analyzing magnet by taking
an effective edge approximation. Matching track segments before and after the analyzing magnets allows for  the  determination of  the track's momentum using the vertical magnetic field, the length traversed in
 the magnetic field region 
 and the orientation of the incoming and outgoing tracks.

Once the momentum is known, the reconstructed tracks are projected 
toward the beam axis and checked for consistency with the collision vertex determined
by the BBCs. 
A $3\sigma$ cut is applied about the mean of the distribution of differences 
between the projected track vertex and the BBC vertex along the beam direction. 
An elliptical cut of $3\sigma$ is applied to the two-dimensional distributions of track intersections with the primary vertex plane. 
This plane is defined as the plane normal to the beam axis that contains the collision vertex.
The rapidity cuts were $|y| < 0.1$ at mid-rapidity and $2.95 < y < 3.15$ at forward rapidity.

\subsection{Particle Identification} 
\begin{figure}[htb]
\includegraphics[width=1.1\columnwidth]{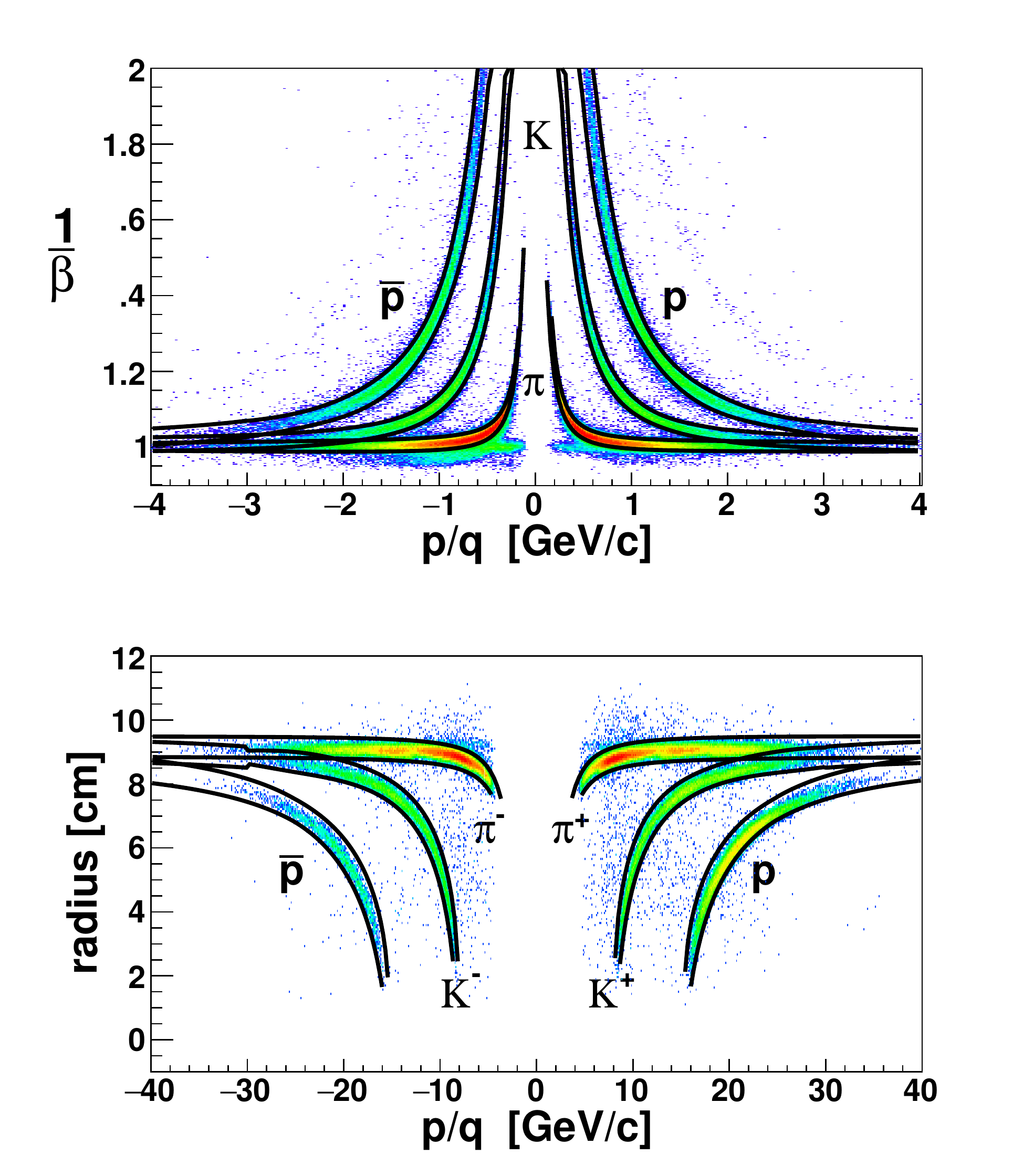}
\caption{(Color online) Scatter plots of $\frac{1}{ \beta}$ versus $p/q$  (top) and RICH radius versus $p/q$ (bottom) for Cu+Cu collisions at $\sqrt{s_{NN}}$ = 200\,GeV. 
The solid curves show the $3\sigma$ cuts around the nominal values given by Eqs. (\ref{EQ1})
 and (\ref{EQ3}). \label{PID}}
\end{figure}

In this analysis, the MRS time of flight and the FS RICH detectors are used for PID at $y=0$  and $y=3$, respectively. 
The time of flight measurement with TOFW and knowledge of the flight  path length allows  $\beta$ to be determined. 
This  together with the  momentum of a detected particle provides for particle identification  using the relation 
\begin{equation}
\frac{1}{\beta^2} = \frac{m^2}{p^2} + 1~.
\label{EQ1}
\end{equation}
Particles of different masses fall on separate curves if $\frac{1}{\beta}$ 
is plotted versus momentum. 
The TOFW provides $\pi/K$ separation up to a momentum of 2~GeV/$c$ and $K/p$ separation up to 3~GeV/$c$.
Figure \ref{PID} (top) shows the distribution of $\frac{1}{\beta}$ $\it{vs.}~p$ for the MRS where $q = 1$ for positive
particles and $q = -1$ for negative particles. 
For this analysis, tracks were required to
have measured $\frac{1}{\beta}$ values within $3\sigma$ of the nominal values
given by Eq.(\ref{EQ1}) for each particle species.  
The curves show the $3\sigma$ cuts around the nominal trajectories for  the different particle species.

For the FS, the emission angle $\theta_c$ of the light radiated in the RICH detector along the particle path is given by
\begin{equation}
\cos\theta_c = \frac{1}{n\beta}, \hspace{10pt} \\
\label{EQ2}
\end{equation}
where $n$ is the index of refraction of the gas inside the RICH volume. 
A spherical mirror of focal length $L$ was used to focus the light cones onto rings of radii
\begin{equation}
r = L \cdot \tan\theta_{c} .
\label{EQ2a}
\end{equation}
Once the radii of the Cherenkov rings are measured, the masses of the particles are deduced from the   formula
\begin{equation}
r = L\tan~\lbrack \cos^{-1}(~\frac{1}{n}\sqrt{1+\frac{m^2}{p^2}}~)~\rbrack. 
\label{EQ3}
\end{equation}
The RICH can identify pions starting at 2.5 GeV/$c$, kaons starting
around 8 GeV/$c$, and protons (anti-protons) from 15\,GeV/$c$. 
The $\pi/K$ $3 \sigma$ separation extends up to 20 GeV/$c$ and protons (anti-protons) can be identified up to 35 GeV/$c$. 
Figure \ref{PID} (bottom) shows
the distribution of radius $r~\it vs.~p$ for the RICH detector. 
At $y=3$, the tracks were required to have a RICH radius within $3\sigma$ of 
the nominal radius for a given species 
as determined from Eq.~(\ref{EQ3}), with a correction to the yield applied for purity in the overlap regions. 

\subsection{Corrections}

The data presented are corrected for the geometrical acceptance of the spectrometers,
tracking efficiency, particle mis-identification and the effects of particle decays based on the GEANT3 simulations. 
These  simulations are also  used to correct the experimental results for effects such as
interactions with the beam pipe, absorption, and multiple 
scattering within the gas volumes of the tracking detectors.

To account for the acceptance,  particles are  generated  with a uniform momentum distribution over a range of angles $\Delta\phi$ and $\Delta\theta$ broad enough for the spectrometer aperture to lie  within the range. 
The acceptance factor for a given pseudo-rapidity and $p_T$ range is then the 
fraction of accepted particles to those thrown scaled by $\frac{\Delta\phi}{2\pi}$. 
This is done for each vertex bin and for the different spectrometer angle and magnetic 
field settings. The acceptance correction is applied to the individual spectra from 
different spectrometer settings before they are averaged.

The tracking efficiency is calculated using a reference track method where good tracks
from one set of detectors are taken as input to a detector whose efficiency is sought.
For the MRS, for example, tracks from the first time projection chamber (TPC) and the Time of Flight wall 
are used as input to determine the efficiency for the second TPC, and vice-versa. 
The ratio of the number of tracks matching the reference tracks to the total number of
input reference tracks is taken as the tracking efficiency. The product of the 
efficiencies calculated for the two MRS TPCs in this way is then taken to be the overall
tracking efficiency for the MRS and is $\sim 92\%$. 
For the FS, the overall tracking
efficiency is $\sim 80\%$, determined as the product of the individual efficiencies for all 
tracking stations. 
The systematic uncertainty on the final spectra
 associated with the determination
of the tracking efficiency 
is $\sim 5 - 8\%$. The tracking efficiency is applied to 
the final MRS spectra.  For the FS the efficiency correction was applied on a track by
track basis.

The corrections for multiple scattering and hadronic absorption were computed by 
 simulating single particle events 
 with  GEANT3 (including
the relevant physical processes in the detector material)
 and processing the results through  the standard BRAHMS analysis code.
The simulations included multiple scattering and 
hadronic interaction processes. These GEANT corrections are applied on a track by track basis
for both the MRS and FS.

To take into account particle mis-identification, a PID correction has been applied 
to the pion and kaon spectra. 
At higher momenta the well defined 3$\sigma$ bands start to overlap. The contamination of the pions and kaons was evaluated by fitting the distributions 
in  m$^2$, $\frac{1}{\beta}$ or ring radius for narrow $p_T$ bins and determining  the contamination fractions and their systematic uncertainties. The invariant yields have been corrected due to this effect.
Typical correction factors are given in Table \ref{tab:PIDCorrection}.

In the momentum range covered, the (anti)protons are well separated from the mesons and no PID correction is applied to their spectra.   
\begin{table}[htdp]
\begin{center}
\begin{tabular}{|c|c|c|c|c|} 
\hline 
 & \multicolumn{2}{|c}{\bf  y=0} & \multicolumn{2}{|c|}{\bf y=3} \\
 &  1.5 GeV/$c$ & 2.25 GeV/$c$ & 24 GeV/$c$ & 30 GeV/$c$ \\ \hline 
Pion & $>99\%$ & $85\%\pm1\% $ &  $>99\%$ & $ 88\% \pm 5\%$ \\ \hline
Kaon   & $> 99\%$ & $ 50\% \pm 5\% $ & $> 99\%$ & $65-70\% \pm 5\%$ \\ \hline
\end{tabular}
\end{center}
\caption{Purity estimates of the  pion and kaon raw spectra, $c_{PID}$,  and their relative systematic uncertainties for pions and kaons at central and forward rapidity for various momenta.As an example the raw pion spectrum at 2.25 GeV/$c$ is corrected by a factor of $0.85 \pm 0.01.$}
\label{tab:PIDCorrection}
\end{table}

Feed down from $\Lambda$-decay corrections are not applied to the proton (anti-proton) spectra.  
This is primarily 
because the spectra of $\Lambda$'s have not been measured at the higher rapidities. 
Later, when discussing integrated yields ($dN/dy$) of protons at mid-rapidity those have been corrected to first order 
since  the $\Lambda$ yields 
were measured by other experiments
\cite{Abelev:2006cs,Agakishiev:2011ar}, 
and detailed simulations indicate that about 90\% of the
decay protons from $\Lambda$s 
pass our cuts for primary particles.

\section{Results and discussion}
\subsection{Particle spectra}

\begin{figure*}[ht]
\includegraphics[width=1.0\textwidth,angle=0]{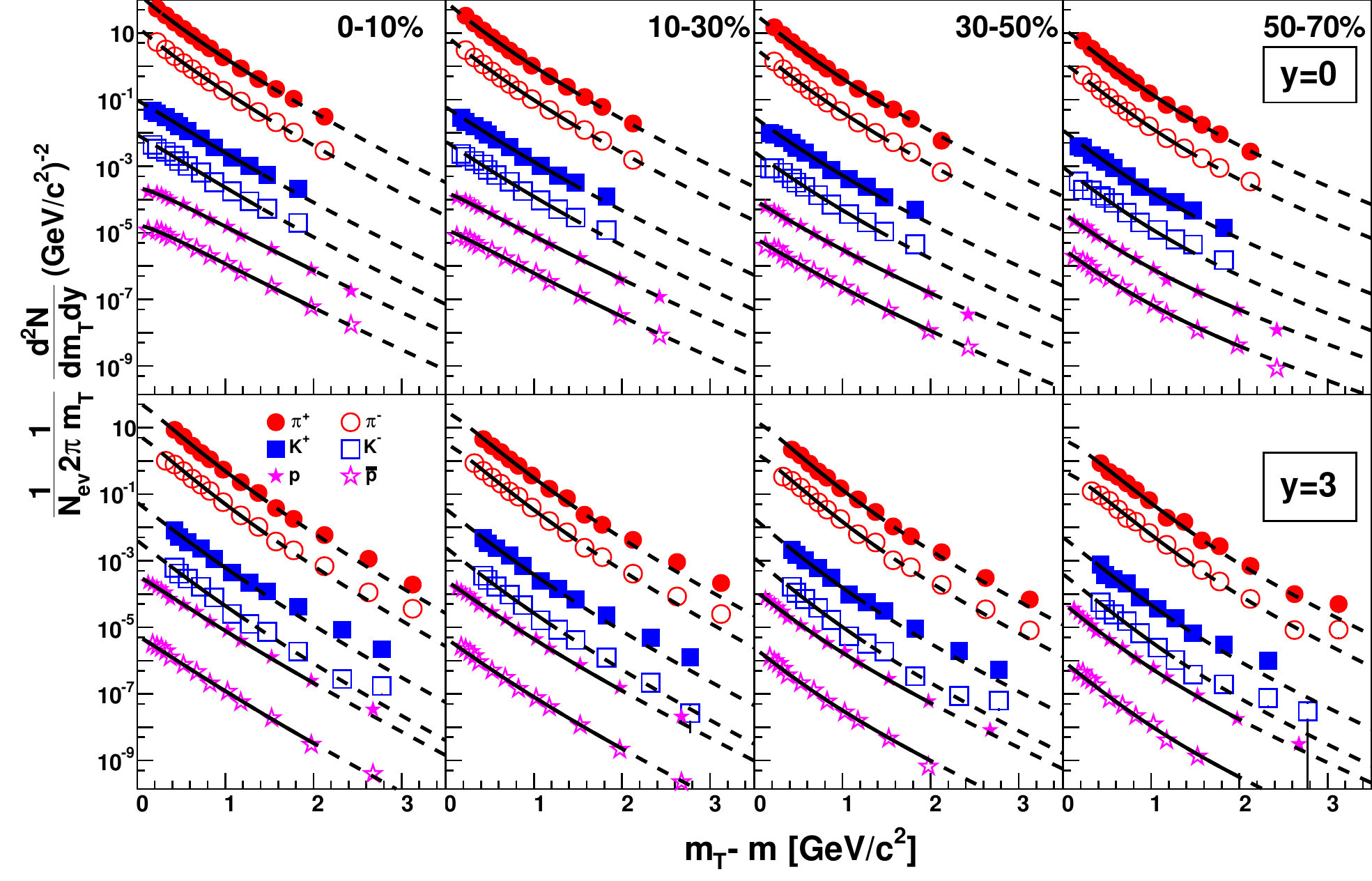}
\caption{(Color online) Invariant spectra from Cu+Cu collisions at $\sqrt{s_{NN}} = 200$ GeV   versus transverse kinetic energy at $y = 0$ (top panels) and $y = 3$ (bottom panels) 
as a function of centrality for $\pi^{\pm},K^{\pm},p$, and $\bar p$.  The $\pi^-,K^+,K^-,p$ and $\bar p$ spectra are scaled by factors of $10^{-1},10^{-2},10^{-3},10^{-4}$, and $10^{-5}$, respectively. 
The lines show the results of blast wave fits to each of the six spectra at a given rapidity and centrality. The solid lines indicate the fit range used while the dashed lines are extrapolations of the functions beyond the fit range. 
Only statistical errors are shown. \label{Spectra}}
\end{figure*}

Measurement of transverse momenta spectra is the crucial first step in obtaining the various observables
 used to characterize the properties of the partonic medium created in heavy ion collisions.
Figure~\ref{Spectra} shows the invariant spectra for the charged hadrons $\pi^{\pm},K^{\pm},p$ and $\bar{p}$, versus transverse kinetic energy, 
for different collision centralities at $ y = 0$ and $y = 3$. 
The spectra of particles and antiparticles have very similar shapes. 
Comparing pions, kaons and protons, a steady hardening of the spectra with particle mass is observed. Both of these effects are suggestive of hydrodynamics. 
The lines in Fig.~\ref{Spectra} are fits of the hydrodynamical inspired blast wave model \cite{Tomasik:1999cq}
 to the six  $\pi^{\pm},K^{\pm},p$ and $\bar{p}$ spectra at a given rapidity and centrality. 
These fits will be discussed in detail later. 
The magnitude of the spectra depend strongly on centrality for all particles and for both rapidities. 
For kaons and protons the shapes of the spectra harden as one moves from peripheral to central collisions. 
The spectra for all particle species are softer at forward rapidity
but, again, one observes a strong centrality dependence.

A systematic study of the spectra was performed by fitting them to a variety of functions. 
For pions  the Levy function 
$A \cdot \Big(1+\frac{(m_T-m_0)}{n_{0}T}\Big)^{-n_0}$
\cite{Wilk:Levy,STAR:Levy,Adare:2010fe}
provided the best fit to the data  
while kaons and protons were best fit by the Boltzmann function  ($A \cdot m_Te^{-\frac{m_T}{T}}$).  
For the   Boltzmann function, the fit parameter $T$ can be  thought of as the effective  temperature of the system. 
The resulting fitting parameters, the $\chi^2$ per number of degrees of freedom, and NDF,   
are listed in  Tables~\ref{table:MrsPions},  \ref{table:MrsKaons}, and \ref{table:MrsProtons}  for pions, kaons, and protons, respectively. 
The integrated yields  $\frac{dN}{dy}$, and mean transverse momenta,  $\langle p_{T} \rangle$,
are obtained by extrapolating the fit functions outside the measurement region. 
The fraction of the particle yield within the BRAHMS acceptance varies from $~30-75\%$ depending upon the spectrometer setting and particle specie. 
Results from other functions were used to estimate the contribution to systematic errors on $\frac{dN}{dy}$ and $\langle p_{T} \rangle$ from the extrapolation beyond the acceptance of the experiment.

A model dependent analysis of the transverse momentum spectra as a function of rapidity 
and centrality allows the extraction of the thermodynamic and collective properties 
of the system at kinetic freeze-out.
At mid-rapidity the hydro-inspired blast wave model \cite{Tomasik:1999cq} 
 predicts a spectrum  with
\begin{equation}
\frac{dN}{m_{T} dm_{T}} \sim \int_{0}^{R_{max}}dr~\lbrace r\times n(r)\times[m_{T} I_{0}(x)K_{1}(z)] \rbrace
\label{Eqn:Blastwave}
\end{equation}
where $x = \frac{p_{T}}{{\rm T}_{\rm kin}} \sinh(\rho)$, $z = \frac{m_{T}}{T_{\rm kin}} \cosh(\rho)$, 
$\rho = \tanh^{-1}(\beta_{T})$, and $\beta_{T}(r) = \beta_{s}(\frac{r}{R})^{\alpha}$ 
is the velocity profile 
as a function of radial distance, $r$. 
In this model T$_{\rm kin}$ represents the kinetic temperature of the system, $\beta_{s}$ the velocity of the surface of the expanding medium and $\alpha$ controls how the velocity of the expanding matter depends upon radial distance. 
For this study $R$ was taken to be the nuclear radius.  
In Eq.~(\ref{Eqn:Blastwave}), $n(r)$ is the radial density profile. 
 In this analysis $n(r)$ is assumed to have a Gaussian form $\sim e^{-\frac{r^2}{2R^{2}}}$ 
 for $r < R_{max}$ where $R_{max}= 3R$. For $r > R_{max},  n(r) = 0.$ 
 The modified Bessel function $K_{1}(z)$ comes from integration from $- \infty$ to $+ \infty$ 
over pseudo-rapidity $\eta$ assuming boost invariance. At forward rapidity, the assumption of 
boost invariance is not valid and $K_{1}(z)$ should be replaced by an integral over 
over a finite range of $\eta$  so that 
\begin{equation}
\frac{dN}{dy m_{T} dm_{T}} \sim \int_{0}^{R_{max}}dr~\lbrace r\times n(r)\times[m_{T} I_{0}(x)g(z)] \rbrace
\label{EQ5}
\end{equation}
where 
\begin{equation}
g(z) = \int_{\eta_{min}}^{\eta_{max}} \cosh(\eta - y)~e^{-z \cosh(\eta - y)}~d\eta
\label{EQ6}
\end{equation}
and $y$ is the rapidity variable. 
The limits of the integration in Eq. (\ref{EQ6}) 
were $\eta_{min}  = 2.4$  and $\eta_{max}  = 4.4$.  
At these limits the integrand in 
 Eq.~(\ref{EQ6}) is very small compared to its central value at $\eta=3$. The results of the fit are stable with respect to small changes in these limits. 

For both the mid-rapidity and forward-rapidity data, we performed a simultaneous fit 
of the pion, kaon and (anti)proton spectra with 3 parameters: $T_{\rm kin}, \beta_s,$ and $\alpha$.
The normalization parameters are adjusted such that the integral yield of the data in the fitting range is reproduced. 
Feed down from resonances was not considered since  the
data do not extend below 0.4 GeV/$c$ where such effects are likely to be significant. 
The fit ranges for pions, kaons and protons are restricted to $p_T < 1.8$\,GeV/$c$ , $p_T < 2.0$\,GeV/$c$, and $p_T < 3.0$\,GeV/$c$,
respectively, since hard processes are expected to become significant above these momenta.

The fits are shown as lines in Fig.~\ref{Spectra}. 
The solid lines indicate the transverse mass range for the fits and the dotted lines are extrapolations of the functions beyond the fit range.  
The systematic errors on the parameters were estimated by changing the fit ranges used for the fits, using different density profiles, and different maximum radii, and for the forward data changing the limits of the $\eta$ integration.
The blast wave fit parameters are tabulated in Table~\ref{table:TempAndBeta}. 
The $\frac{dN}{dy}$ and  $\langle p_{T} \rangle$ from the blast wave fit are in reasonable agreement with fits to the individual  kaon and proton spectra listed in Tables~\ref{table:MrsKaons} and \ref{table:MrsProtons}.

\begin{figure}[h]
\includegraphics[width=\columnwidth]{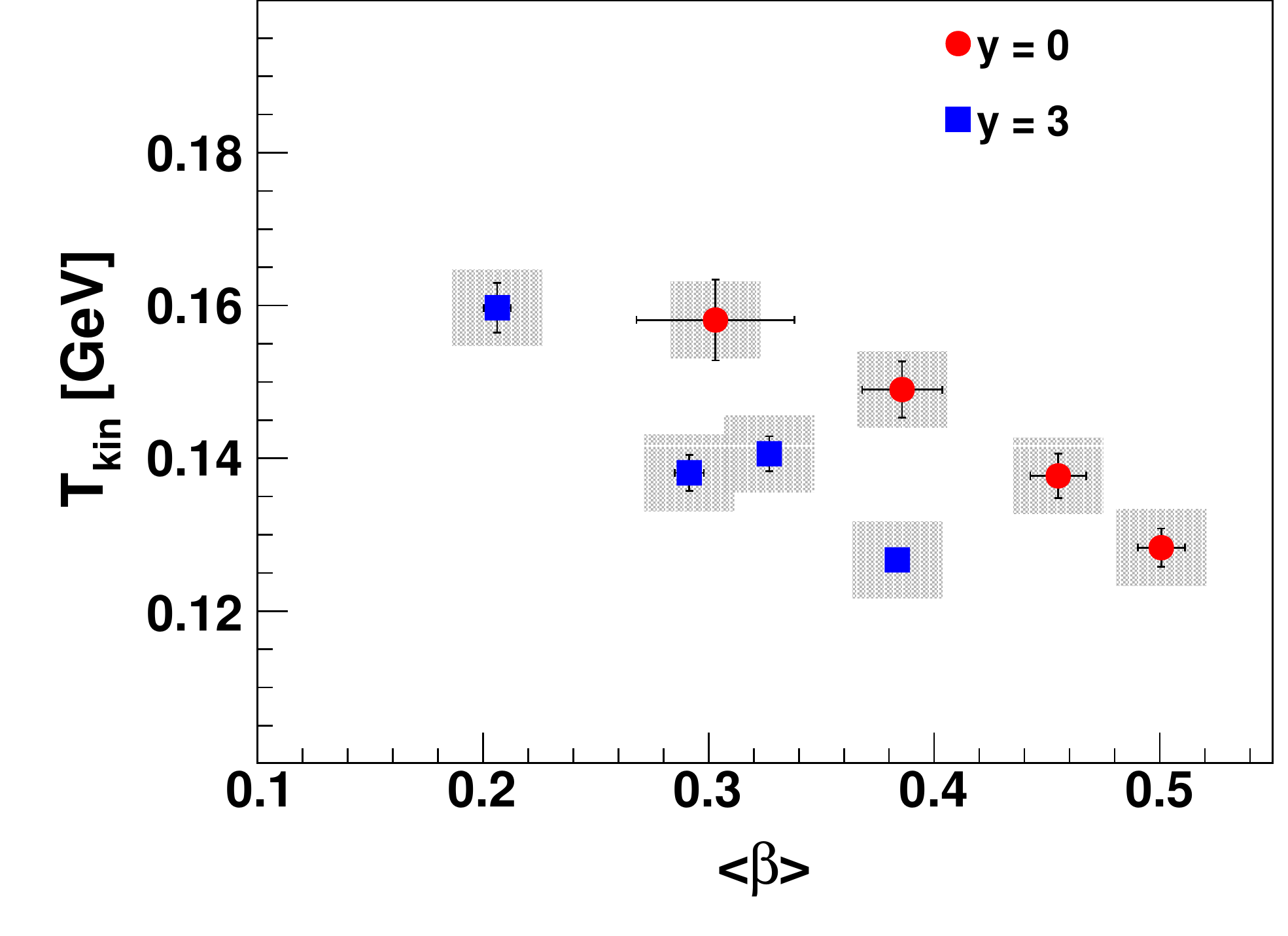}
\caption{(Color online) Blast wave fit parameters T$_{\rm kin}$ vs. $\langle  \beta \rangle$ for 
Cu+Cu collisions at $\sqrt{s_{NN}} = 200$ GeV at $y=0$  (red circles) and $y=3$ (blue squares). 
The statistical errors are represented by bars and the systematic errors  by the gray boxes.  More central collisions are to the right.
The numerical values are listed in 
Table~\ref{table:TempAndBeta}.
\label{TempVsBeta}}
\end{figure}

 Figure~\ref{TempVsBeta}  shows the (anti-)correlation between the kinetic temperature,  T$_{\rm kin}$,  and the average transverse velocity $\langle  \beta \rangle = \frac{2}{\alpha +2} \cdot \beta_s $,   for the four centrality classes at both rapidities. 
 As collisions become more central (going from left to right in Fig. 3), T$_{\rm kin}$ decreases as $\langle  \beta \rangle$
increases.
 This is expected since a   larger system should stay together for a longer time. 
 As the system cools random thermal motion of the partons is converted to bulk radial flow, lowering the temperature and increasing the average velocity.

\begin{figure}[h]
\includegraphics[width=0.9\columnwidth,angle=0]{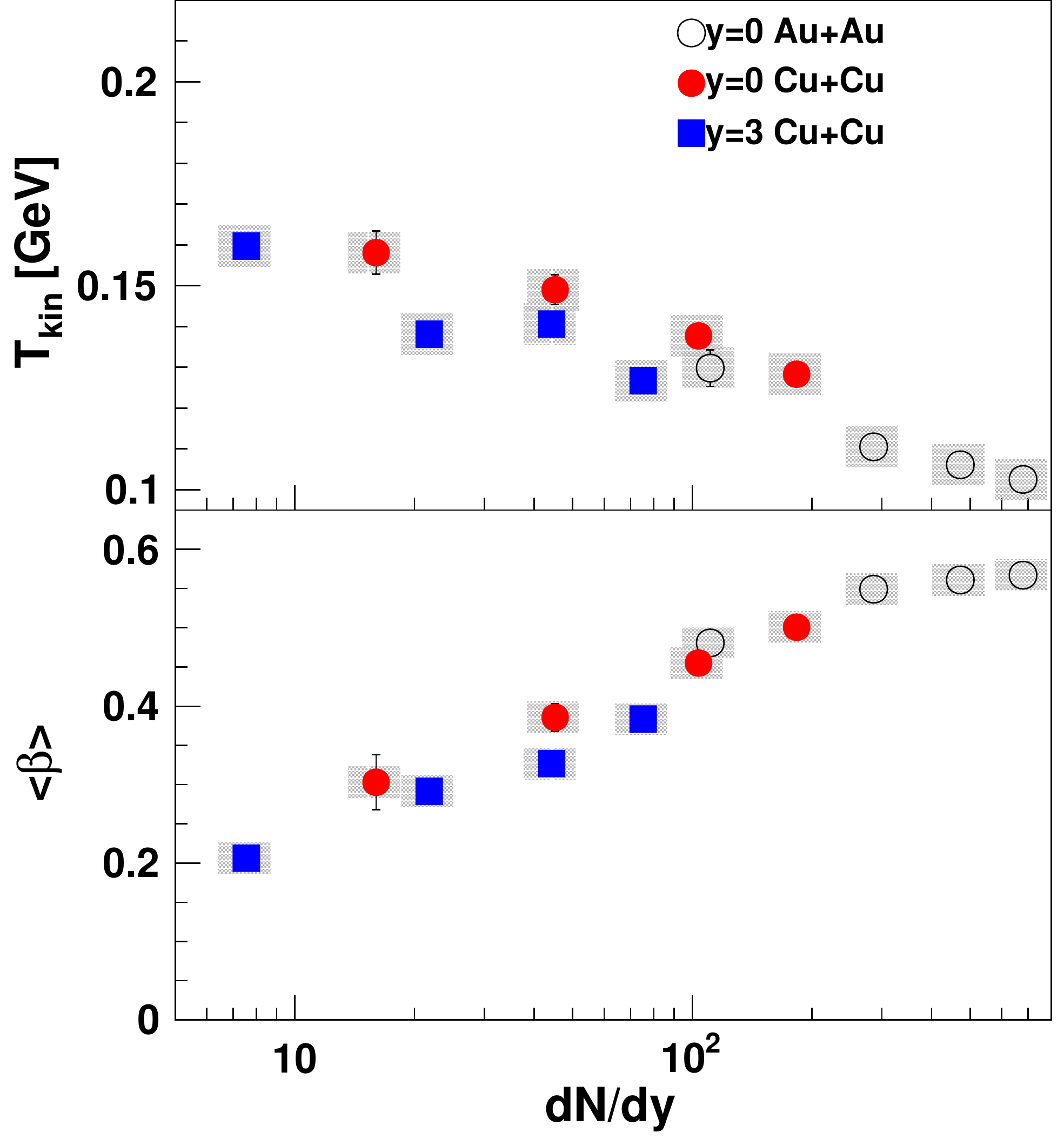}
\caption{(Color online) \label{TBvsCent} T$_{\rm kin}$ (top) and $\langle \beta \rangle$ (bottom) for 200 GeV Au+Au collisions at $y=0$  and Cu+Cu collisions at $y=0$  and $y=3$ as a function of total $\frac{dN}{dy}$ ($\pi^\pm, K^\pm, p$ and ${\bar p}$) for various centralities.
The statistical errors are represented by bars and the systematic errors  by the gray boxes. 
The Au+Au spectra used for the fits  are from \cite{Arsene:2005mr}. 
}
\end{figure}

At $y=3$ the slope of the T$_{\rm kin}$ versus $\langle  \beta \rangle$ curve is similar to that at $y=0$,  but for a given $\langle  \beta \rangle$ the temperatures are  about 20 MeV lower. 
This effect does not just result from having lower particle densities at $y=3$.   
Figure \ref{TBvsCent} shows the  dependence of the kinetic freeze-out temperature and the mean
radial flow velocity for Cu+Cu and Au+Au collisions as a function of the total $dN/dy$ ($\pi^\pm, K^\pm, p$ and ${\bar p}$) of each centrality class at a given rapidity.  
For a given $dN/dy$ both T$_{\rm kin}$ and $\langle  \beta \rangle$ are smaller at $y=3$   reflecting the lower energy (and hence lower $\langle p_{T} \rangle$) that is available to the matter at forward rapidity. 
At mid-rapidity the dependence of T$_{\rm kin}$  and $\langle  \beta \rangle$ on  $dN/dy$ is similar in  Cu+Cu and Au+Au reactions,  
with slightly higher values of  T$_{\rm kin}$ and slightly lower values of $\langle  \beta \rangle$  in Cu+Cu compared to Au+Au reactions. 
At mid-rapidity the STAR collaboration has made blast wave fits to 
$\pi^\pm, K^\pm$, proton and antiproton spectra \cite{Aggarwal:2010pj}. The  reported  
values for T$_{\rm kin}$ are slightly lower  but consistent within errors to the corresponding BRAHMS results. 
The $p_T$ ranges for the data and fits were also slightly different.

\begin{table}[h]
\small
\begin{tabular}{|c|c|c| l| c|c|}
\multicolumn{5}{c}{{\bf $y=0$  }}\\ \hline
${\bf Cent.}$ & ${\bf T_{\rm kin} (MeV)} $ & $ \boldsymbol{\langle \beta \rangle}$ & $\boldsymbol{~~~~~~~\alpha}$ & $\mathbf{\chi^2/dof}$\\ \hline
$~~0-10\% $ & $128 \pm 3 $ & $0.501 \pm 0.010$ & $0.499 \pm 0.024$ & $0.84$ \\ \hline
$ 10-30\% $ & $138 \pm 3 $ & $0.455 \pm 0.012$ & $0.604 \pm 0.028$ & $1.00$ \\ \hline
$ 30-50\% $ & $149 \pm 4 $ & $0.386 \pm 0.018$ & $0.794 \pm 0.045$ & $1.15$ \\ \hline
$ 50-70\% $ & $158 \pm 5 $ & $0.303 \pm 0.035$ & $1.16   \pm 0.11$   & $2.63$ \\ \hline
\multicolumn{5}{c}{{\bf $y=3$  }}\\ \hline
${\bf Cent.}$ & ${\bf T_{\rm kin} (MeV)} $ & $ \boldsymbol{\langle \beta \rangle} $ & $\boldsymbol{~~~~~~\alpha}$ & $\mathbf{\chi^2/dof}$\\ \hline
$ ~~0-10\% $ & $127 \pm 1 $ & $0.384 \pm 0.004$ & $0.723 \pm 0.011$ & $1.35$ \\ \hline
$ 10-30\% $ & $141 \pm 2 $ & $0.327 \pm 0.003$ & $0.886 \pm 0.014$ & $1.61$ \\ \hline
$ 30-50\% $ & $138 \pm 2 $ & $0.291 \pm 0.007$ & $1.09 \pm 0.02$ & $1.22$ \\ \hline
$ 50-70\% $ & $160 \pm 3 $ & $0.206 \pm 0.006$ & $1.53 \pm 0.03$ & $1.27$ \\ \hline
\end{tabular}\\
\caption{Blast wave fit parameters, 
for  Cu+Cu collisions at $\sqrt{s_{NN}} = 200 $ GeV at various centralities at $y=0$ (top) and $y=3$ (bottom). 
The errors listed are statistical
only. The systematic errors are of the order of 5 MeV for T$_{\rm kin}$ and 0.015 for $\langle \beta \rangle$.}
\label{table:TempAndBeta}
\end{table}

\begin{figure*}[ht]
\centering
\begin{minipage}[h!]{1.0\textwidth}
\centering
\includegraphics[width=1.0\textwidth,angle=0]{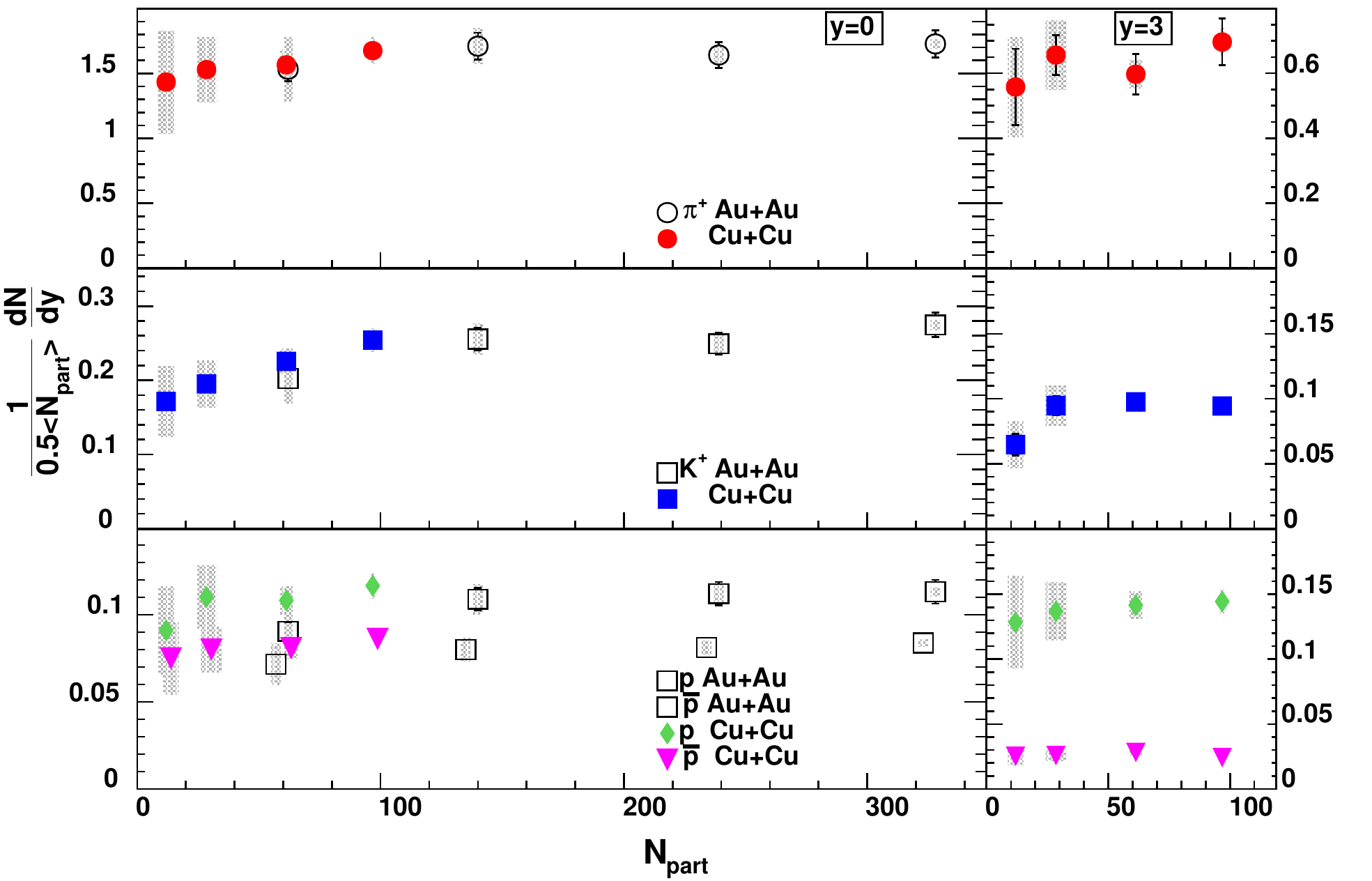}
\end{minipage}\hfill
\caption{(Color online) N$_{\rm part}$  scaled $\frac{dN}{dy}$ for $\pi^+, K^+$ and proton and anti-protons
from Cu+Cu (solid symbols) and Au+Au  collisions (open symbols) at $\sqrt{s_{NN}} = 200 $ GeV
as a function of N$_{\rm part}$  for at $y=0$   (left) and  $y=3$   (right). 
The Au+Au data are from \cite{Arsene:2005mr}. 
The statistical errors are represented by bars and the systematic errors  by the gray boxes. 
The Au+Au pion yields were deduced using a power law extrapolation at low $p_T$.
 \label{ScaledDnDy}}
\end{figure*}

\begin{figure}[ht]
\includegraphics[width=1.0\columnwidth,angle=0]{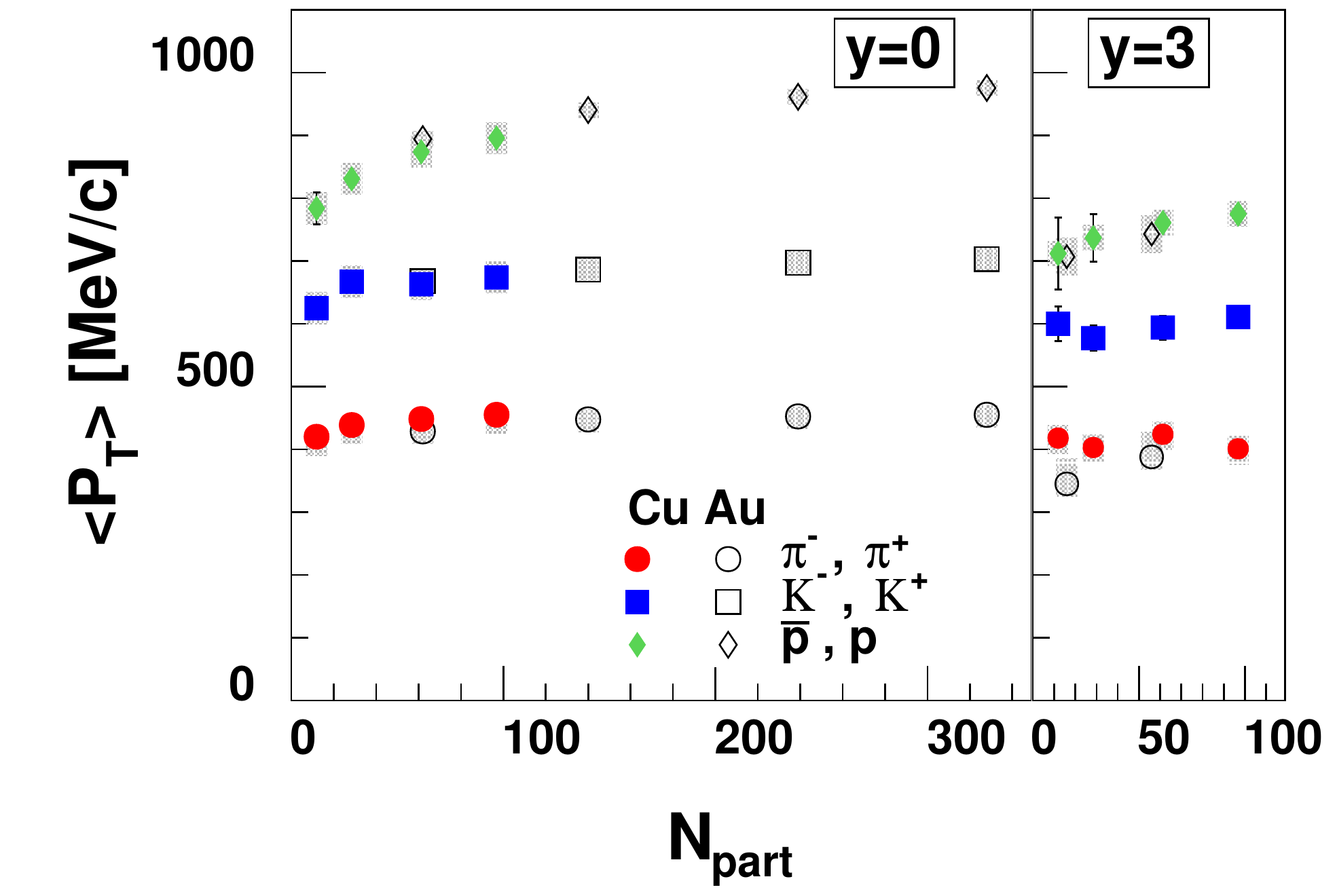}
\caption{(Color online) Mean transverse momentum $\langle p_T \rangle$ for $\pi^\pm, K^\pm$, p and ${\bar p}$ as a function of N$_{\rm part}$  for Cu+Cu and Au+Au collisions at $\sqrt{s_{NN}} = 200 $\,GeV for $y=0$   (left) and $y=3$   (right). 
The Au+Au mid-rapidity data are from \cite{Arsene:2005mr}, and the forward Au+Au proton and pion preliminary data from\cite{Videbaek:2009zy}.
\label{MeanPt}}
\end{figure}

In Fig.~\ref{ScaledDnDy} the $dN/dy$  values per participant pair 
are shown for central (left) and forward (right) rapidity for Cu+Cu and Au+Au (mid-rapidity only) collisions \cite{Arsene:2005mr}. 
For clarity, only the positive pions and kaons are shown, but the trends are very similar for the corresponding negative particles.  At both central and forward rapidity the  kaon yields per participant pair 
 are somewhat smaller for lower values of N$_{\rm part}$. 
For N$_{\rm part}$  between  60  and 100 the  $\frac{1}{0.5 N_{part}}\frac{dN}{dy}$  values for $\pi^\pm$  extracted from Cu+Cu collisions are similar to the ones extracted from Au+Au, while for $K^\pm$ the 
 scaled $\frac{dN}{dy}$ values are slightly higher. 
 A similar effect has been seen by STAR where the $K^-$ and $K^0_S$ yields at a given N$_{\rm part}$ are somewhat higher for Cu+Cu than for Au+Au \cite{Aggarwal:2010pj,Agakishiev:2011ar}.

Beccattini and Manninen  have proposed that 
an increase of the scaled  $\frac{dN}{dy}$  values as observed for the kaon yields might reflect
the effect of two sources,  a chemically equilibrated and dense ``core" and a ``corona" of independent nucleon-nucleon collisions 
\cite{Becattini:2008ya}.  
As the centrality of the system decreases the ratio of core to corona changes causing a change in the kaon yield per participant pair. 

Figure~\ref{MeanPt} shows the average transverse momenta 
$\langle p_{T} \rangle$
for pions, kaons and (anti)protons 
 versus N$_{\rm part}$  for Cu+Cu collisions at $y=0$    and $y=3$   and for Au+Au collisions at $y=0$ \cite{Arsene:2005mr}. 
A general observation
is that $\langle p_T \rangle$ depends strongly on particle mass,
reflecting the larger boost given to the heavier particles by radial
flow (as expected from the blast wave model).
While the pion $\langle p_{T} \rangle$ values at $y=3$   are similar to those at $y=0$  , the kaons and (anti)protons
exhibit smaller values at forward rapidity.  This drop in $\langle p_{T} \rangle$ for the heavier particles reflects the lower radial flow and freeze-out temperatures at forward rapidity shown in Figs.~\ref{TempVsBeta} and \ref{TBvsCent}.

At mid-rapidity there is a small increase in the  pion $\langle p_{T} \rangle$ as the collisions become more central.
The increase of the $\langle p_{T} \rangle$ values for more central collisions is more pronounced for  the kaons and the (anti)protons. 
The pions and kaons show no $\langle p_{T} \rangle$-dependence on centrality at
forward rapidity while the (anti)proton $\langle p_{T} \rangle$ appears to increase as the collisions become more central.
 The Cu+Cu data points
 join smoothly with those from Au+Au collisions (a similar result was observed by STAR for the $K^{*^{0}}$ in the two colliding systems \cite{Aggarwal:2010mt}.) 
 This suggests that the  $\langle p_{T} \rangle$ values are insensitive to the difference in shape 
of the Cu+Cu and Au+Au overlap regions for the same number of participants.

\subsection{Nuclear Modification Factors}

\begin{figure*}[ht]
\centering
\begin{minipage}[h!]{1.0\textwidth}
\centering
\includegraphics[width=0.85\textwidth,angle=0]{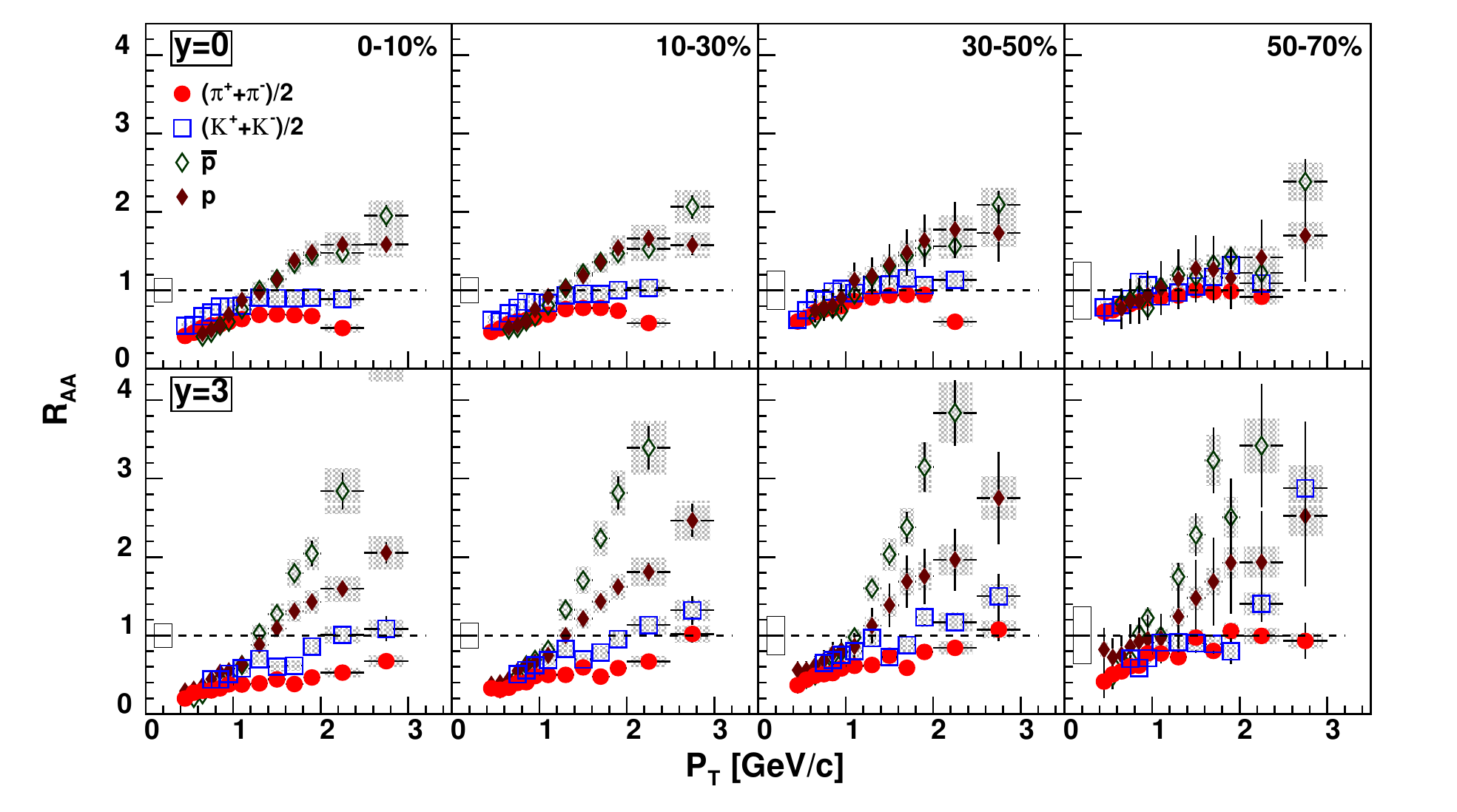}
\end{minipage}\hfill
\caption{\label{RaaVsPt} (Color online) Nuclear modification factor of $\sqrt{s_{NN}} = 200$ GeV Cu+Cu collisions for  pions, kaons and (anti)protons as a function of $p_T$ and centrality. 
The top row is for $y=0$ and the bottom row $y \approx 3$.  
The centrality decreases from left to right.
The statistical errors are represented by bars and the systematic errors  by the gray boxes. The white boxes at $p_T = 0$ represent the correlated normalization error from the $p+p$ reference and the error on the number of participants.  
}
\end{figure*}

The discovery of high $p_T$ hadron suppression at central rapidity in $\sqrt{s_{NN}} = 130$ GeV  Au+Au  collisions at RHIC has been one of the most  exciting results   in heavy ion physics \cite{Adcox:2001jp,Adler:2002xw,Arsene:2003yk,Back:2003qr}.
These first measurements have since been extended to higher energies and a broad range of colliding systems, both light and heavy, and also refined to include identified hadrons, heavy quarks and fully reconstructed jets 
\cite{Adler:2003qi,Adams:2003kv,Adare:2006nq,Aad:2010bu,Chatrchyan:2011sx,Aad:2012vca,Chatrchyan:2012gw}.  
No such effects have been seen at $y \sim 0$  in d-Au collisions at  RHIC  
\cite{Adler:2003ii,Adams:2003im,Arsene:2003yk,Back:2003qr}
 confirming that the observed suppression found 
at mid-rapidity in central heavy-ion collisions is indeed a final-state
effect and is specifically a consequence of the energy loss of partons. 
At forward rapidity, the colliding systems
d+Au and Au+Au at $\sqrt{s_{NN}}$ both exhibit high $p_T$ suppression similar
to each other and to the mid-rapidity Au+Au results \cite{Arsene:2004ux,Arsene:2004ux}.

The nuclear effects  on particle production  are studied in terms of the nuclear modification factor $R_{AA}$ defined as
\begin{equation}
R_{AA} = \frac{d^{2}N_{AA}/dp_T dy}{\langle {\rm N}_{\rm coll}\rangle d^{2}N_{pp}/dp_T dy},
\label{Eqn:Raa}
\end{equation}
which is the ratio of the particle yield in heavy ion collisions to the yield in $p+p$ 
collisions scaled by the average number of binary nucleon-nucleon collisions $\langle {\rm N}_{\rm coll} \rangle$ for a given centrality class.  
 If AA collisions were just a superposition of elementary collisions between nucleons, 
then $R_{AA}$ should be 1.0 in the $p_T$ region dominated by hard processes.  

Partonic energy loss in a hot QGP will typically lead to an $R_{AA}$ value
well below unity.  Initial-state effects, such as shadowing of the nuclear  parton distribution functions  may  also influence the $R_{AA}$ values and are believed to contribute
to the suppression observed at forward rapidity in $\sqrt{s_{NN}}$ = 
200 GeV d+Au and Au+Au collisions at RHIC \cite{Arsene:2004ux,Arsene:2006ts}.  
The particle species 
dependences of $R_{AA}$ at low to intermediate $p_T$ may be influenced 
by various medium effects such as
 collective radial flow (leading to a mass ordering 
of the $R_{AA}$ of identified hadrons) and/or parton recombination 
effects (typically leading to meson-baryon differences).

Figure~\ref{RaaVsPt} shows the nuclear modification factor $R_{AA}$ for pions, kaons, and (anti)protons,
 respectively, in Cu+Cu collisions. 
 The pion and kaon $R_{AA}$ values are averages of the positive and negative particles.
A general trend immediately seen is the clear mass ordering of the
$R_{AA}$ values for the various particle species, most pronounced in the
more central collisions and compatible with radial flow and/or
recombination effects influencing the modification pattern.

For $1 < p_{T}< 2$\,GeV/$c$  the pions are suppressed at both rapidities for central and mid-central events.  
The level of suppression is strongest for more central collisions which achieve the highest densities and largest volumes.
 This is consistent with the fact that the multiplicity density decreases as one goes to more peripheral collisions;
 there is less matter to interact with and more partons make it out of the collision region before losing much of their
 energy. Interestingly the suppression is stronger at forward rapidities where one would expect parton energy loss to be less.  This is consistent with the pattern seen for $\pi^-$ mesons in Au+Au collisions \cite{Arsene:2006ts}.

 Kaons with  $1<p_T<2$ GeV/c do not show significant suppression at $y=0$ but they are suppressed at $y=3$. 
 The suppression of the kaons is less pronounced than that of the pions but shows 
  a similar 
dependence on centrality.  
The difference in the pion and kaon suppression patterns may reveal information about 
their respective  fragmentation functions \cite{Djordjevic:2013qba}. 
At mid-rapidity, the $R_{AA}$ values  for pions and kaons vary little with $p_T$ over the range 
$p_{T} = 1.5 -2.5$\,GeV/$c$. 
At forward rapidity there is an increase of the kaon and pion $R_{AA}$ values with $p_T$
similar for all centralities but somewhat less pronounced for the for  most  peripheral sample.

For both protons and antiprotons $R_{AA}$ rises steadily with $p_T$ crossing 1.0  at $p_T \approx 1.3$\,GeV/$c$ for all centralities and both rapidities. 
The enhancement for $p_T > 1.3$\,GeV/$c$ is strongest in peripheral collisions and at forward rapidity. 
At central rapidity the enhancement is similar for protons and antiprotons but at $y=3$ the antiprotons show a greater enhancement than the protons, partially due to the isospin-related difference in
reference spectra for protons and antiprotons at forward rapidity
in $p+p$ collisions \cite{PhysRevLett.98.252001}.

Note that the STAR collaboration has measured $R_{AA}$ for pions and 
$p+{\bar{p}}$ at mid-rapidity and $p_T \ge 3$\,GeV/$c$ in Cu+Cu collisions at $\sqrt{s_{NN}} = 200$~GeV/$c$ \cite{Abelev:2009ab}.  
The STAR results are consistent with our highest $p_T$ data points but fall steadily  before leveling off at $p_T \approx 6$\,GeV/$c$. 

\begin{figure}[ht]
\includegraphics[width=1.\columnwidth,angle=0]{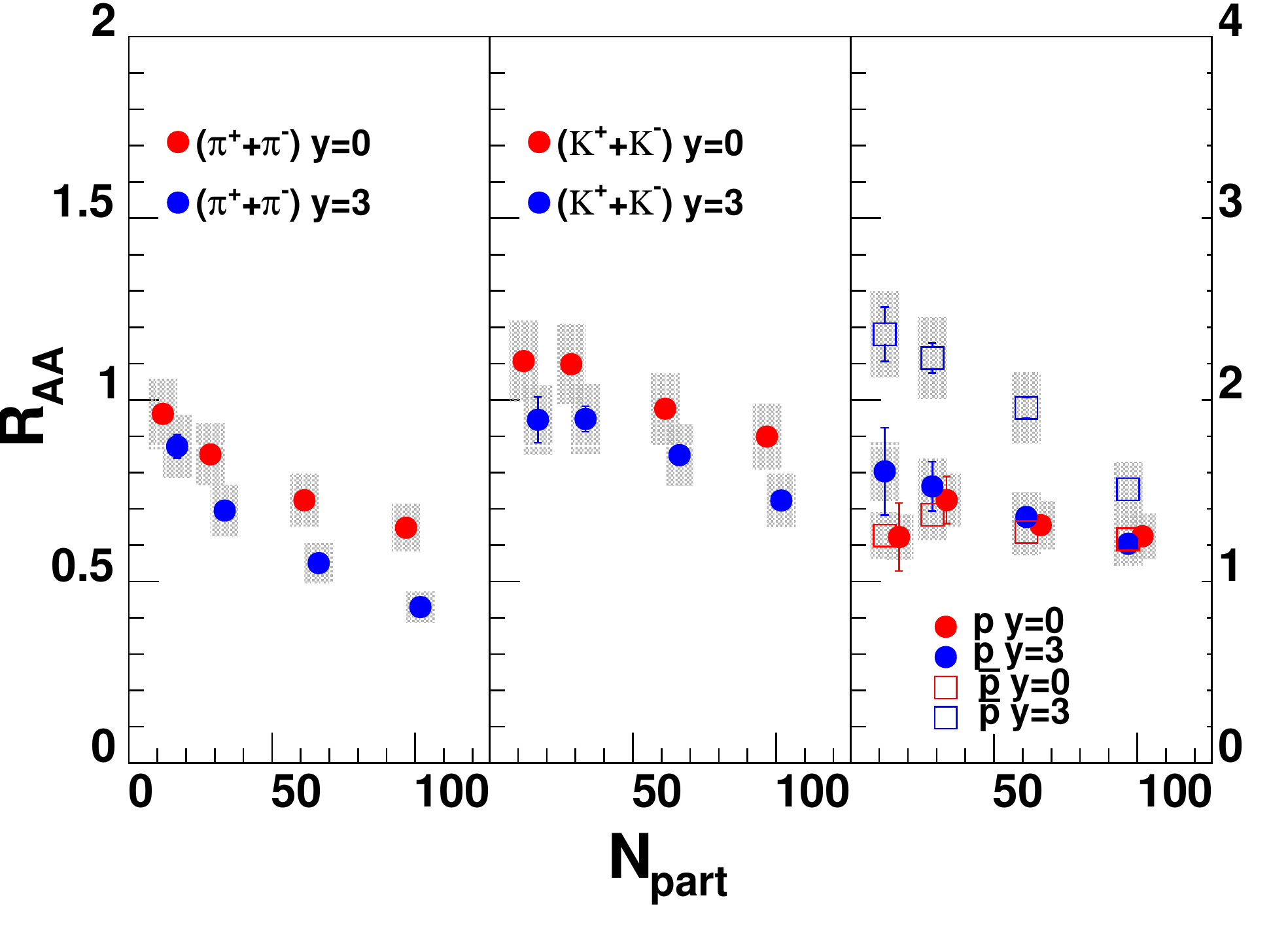}
\caption{(Color online) 
Nuclear modification factor of $\sqrt{s_{NN}} = 200$ GeV Cu+Cu collisions for the $p_T$ region 1.3--2.5 GeV/$c$ 
 for  pions (left), kaons (center)  and (anti)protons (right) as a function of $N_{\rm part}$ for $y=0$  (red symbols) and $y=3$ (blue symbols). Systematic errors are shown by the gray bands. However 
the systematic errors that arise from uncertainties in $N_{\rm coll}$  are common to the $y=0$  and $y \approx 3$ datasets are not included.  These errors are listed in Tab.~\ref{tab:npart}.  
 Note the different vertical scale for the (anti)proton $R_{AA}$.
\label{RaaVsNpart}}
\end{figure}

To improve the statistical precision of the results the Cu+Cu and $p+p$ spectra were summed over 
the $p_T$ region 1.3--2.5\,GeV/$c$  
and the $\pi^\pm$ and $K^\pm$ spectra averaged before taking the ratio shown in Eq.~\ref{Eqn:Raa}. 
The resulting  $R_{AA}$ values as a function of $N_{\rm part}$ are shown in Fig.~\ref{RaaVsNpart}.   
For pions and kaons the $R_{AA}$ values are  smaller at $y=3$ than at $y=0$,
for protons they are similar at the two rapidities,  while for antiprotons the  $R_{AA}$ values are  larger at $y=3$ than at $y=0$.   
For pions and kaons $R_{AA}$ drops with  $N_{\rm part}$  at both $y=0$  and $y=3$, 
while this trend is less clear for the baryons.
For both protons and anti-protons $R_{AA}$ is above 1.0 for all values
of N$_{\rm part}$  and at both rapidities, with anti-protons at $y=3$
standing out as most enhanced and with $R_{AA}$ falling with N$_{\rm part}$ .

The fact that the mesons are more strongly suppressed for more central collisions is expected from  from models of parton energy loss or jet quenching. In such models it is expected that the energy loss should be less at forward rapidities because of the decreasing particle density. However this effect my be compensated in the $R_{AA}$ ratio by a relative softening of the Cu+Cu $p_T$ spectra at forward rapidities.  PHENIX has suggested that a similar effect may explain why at high $p_T$ $R_{AA}$ is almost the same at 
 $\sqrt{s_{NN}} = 63$ and 200 GeV \cite{PhysRevLett.109.152301}.
 It is also possible that at forward rapidity initial state effects such as nuclear shadowing are reducing particle production,  \cite{Arsene:2004ux,Arsene:2006ts}.

\subsection{Particle Ratios}
\begin{figure}[ht]
\includegraphics[width=1.0\columnwidth,angle=0]{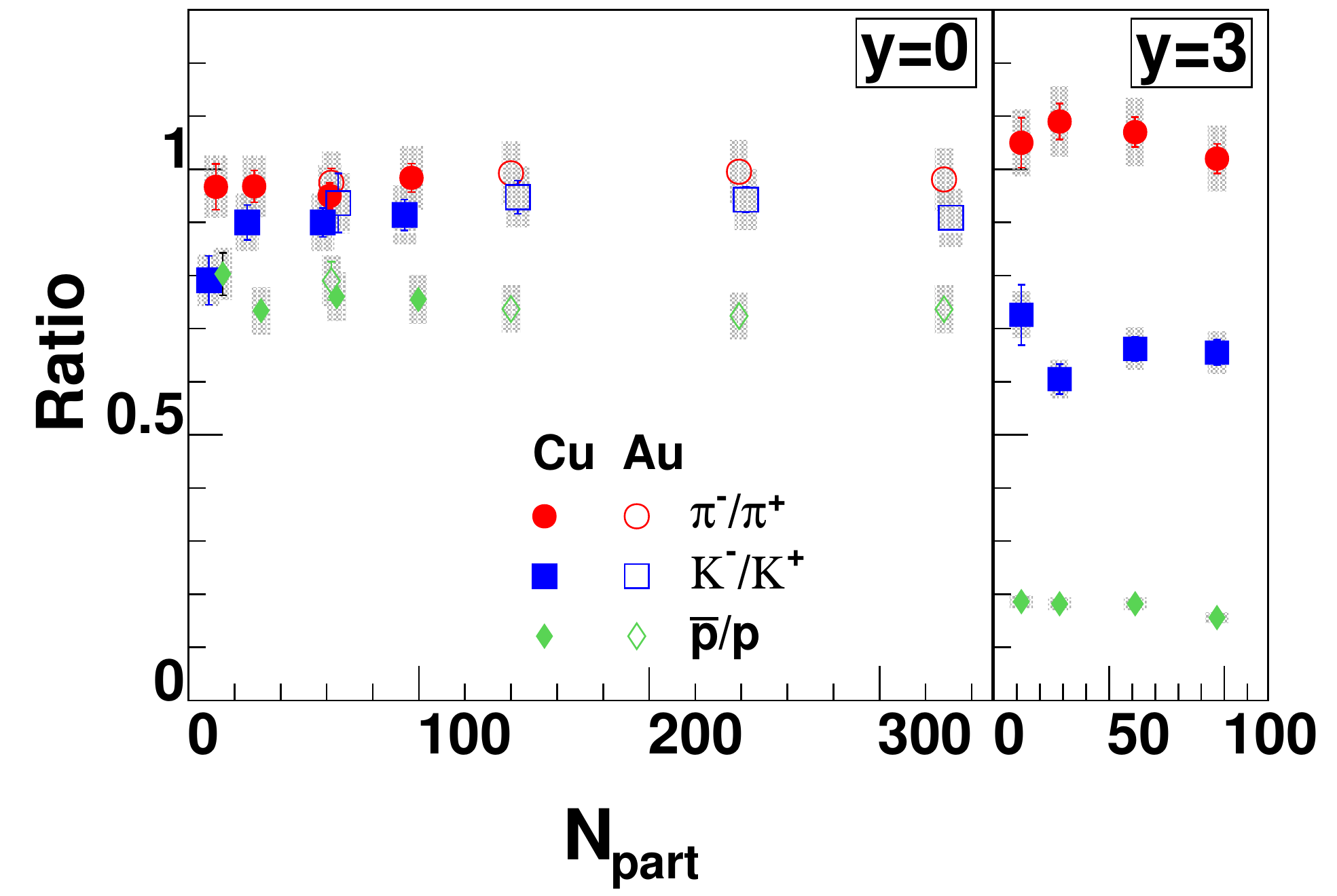}
\caption{(Color online) Ratios of antiparticle/particle yields versus N$_{\rm part}$  for pions, kaons and protons at $y=0$   (left) for  Cu+Cu (solid symbols) and Au+Au (open symbols) collisions  and (right) for Cu+Cu collisions at $y \approx 3$. Both Cu+Cu and Au+Au collisions are at $\sqrt{s_{NN}} = 200$\,GeV. The statistical errors are represented by bars and the systematic errors  by the gray boxes.  \label{RatioVsNpartNew}}
\end{figure}

Figure~\ref{RatioVsNpartNew} shows antiparticle to particle $\frac{dN}{dy}$ ratios of integrated yields measured in Cu+Cu and Au+Au collisions at $\sqrt{s_{NN}} = 200$\,GeV 
as a function of  N$_{\rm part}$, for $y=0$ and $y~\sim~3$.  
These ratios of integrated yields do not exhibit
 a centrality dependence at mid-rapidity. There is very little difference between the Cu+Cu and Au+Au results.  At  $y~\sim~3$ there is a slight drop of the $\frac{\pi^-}{\pi^+}$ ratio with N$_{\rm part}$. 

\begin{figure*}[ht]
\includegraphics[width=1.0\textwidth,angle=0]{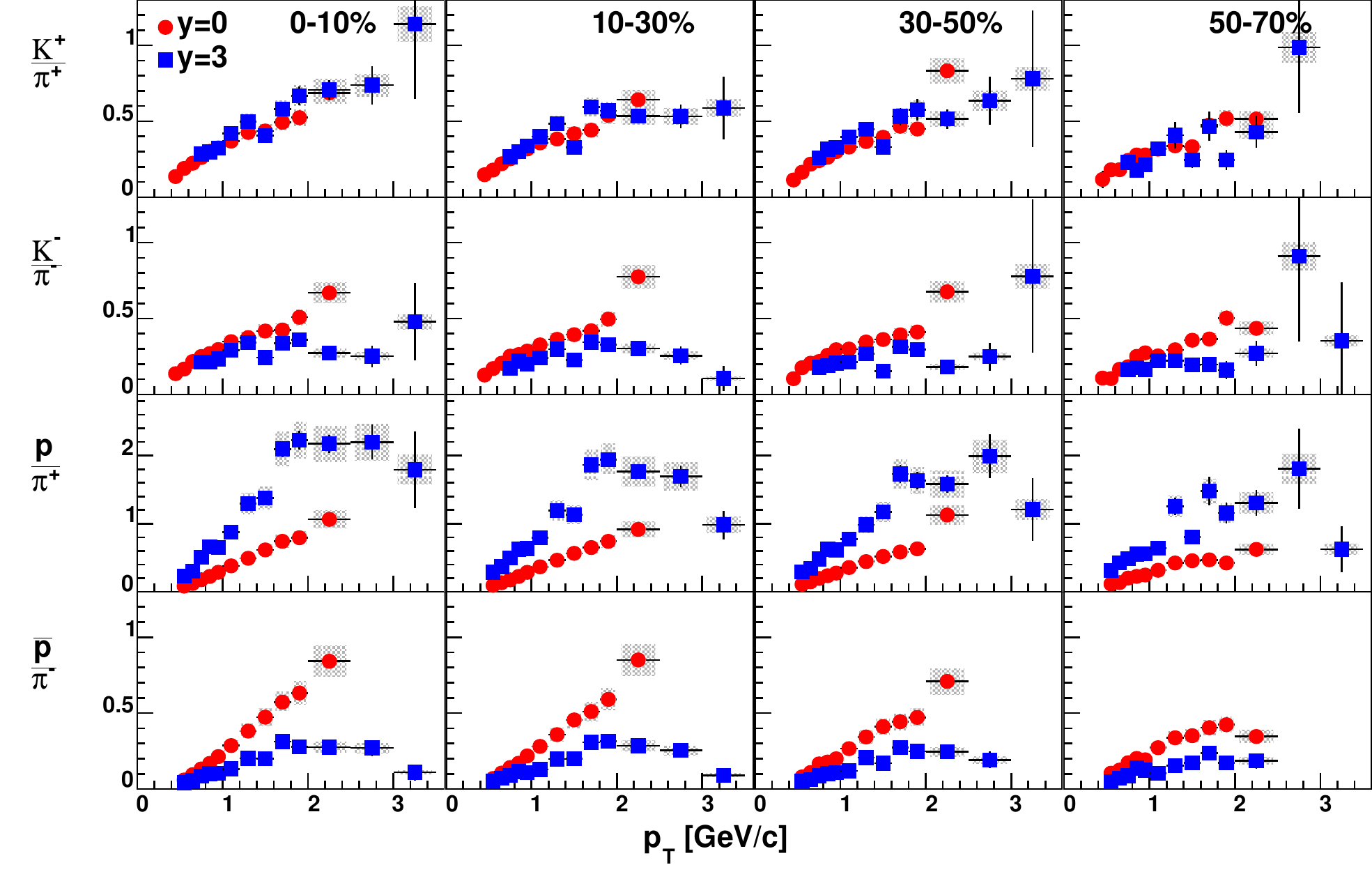}
\caption{(Color online) Particle  ratios from Cu+Cu collisions at $\sqrt{s_{NN}} = 200$ GeV   as a function of $p_{T} $, at $y = 0$ (red circles) and $y = 3$ (blue squares) for various centralities. 
The centrality decreases from left to right.  
The statistical errors are shown by bars and the systematic errors  by the gray boxes. 
\label{ParticleRatioVsPt}}
\end{figure*}

Figure  \ref{ParticleRatioVsPt} shows the kaon to pion ratios (upper two panels) and proton to meson ratios 
(lower two panels) as functions of $p_T$, centrality and rapidity.
At mid-rapidity, the  $\frac{K}{\pi}$ ratios show a linear increase at low $p_T$ but increase less rapidly   
for $p_T > 1.5$~GeV/$c$,  with the $\frac{K^{+}}{\pi^{+}}$ ratio showing only a slight
 excess over the corresponding $\frac{K^{-}}{\pi^{-}}$ values. 
 At $y=3$, both $\frac{K}{\pi}$ ratios show a slightly stronger
centrality dependence than at $y=0$, and the $K^+/\pi^+$ ratios are
significantly enhanced over the corresponding $K^-/\pi^-$ results.
 
\begin{figure}[]
\includegraphics[width=1.0\columnwidth,angle=0]{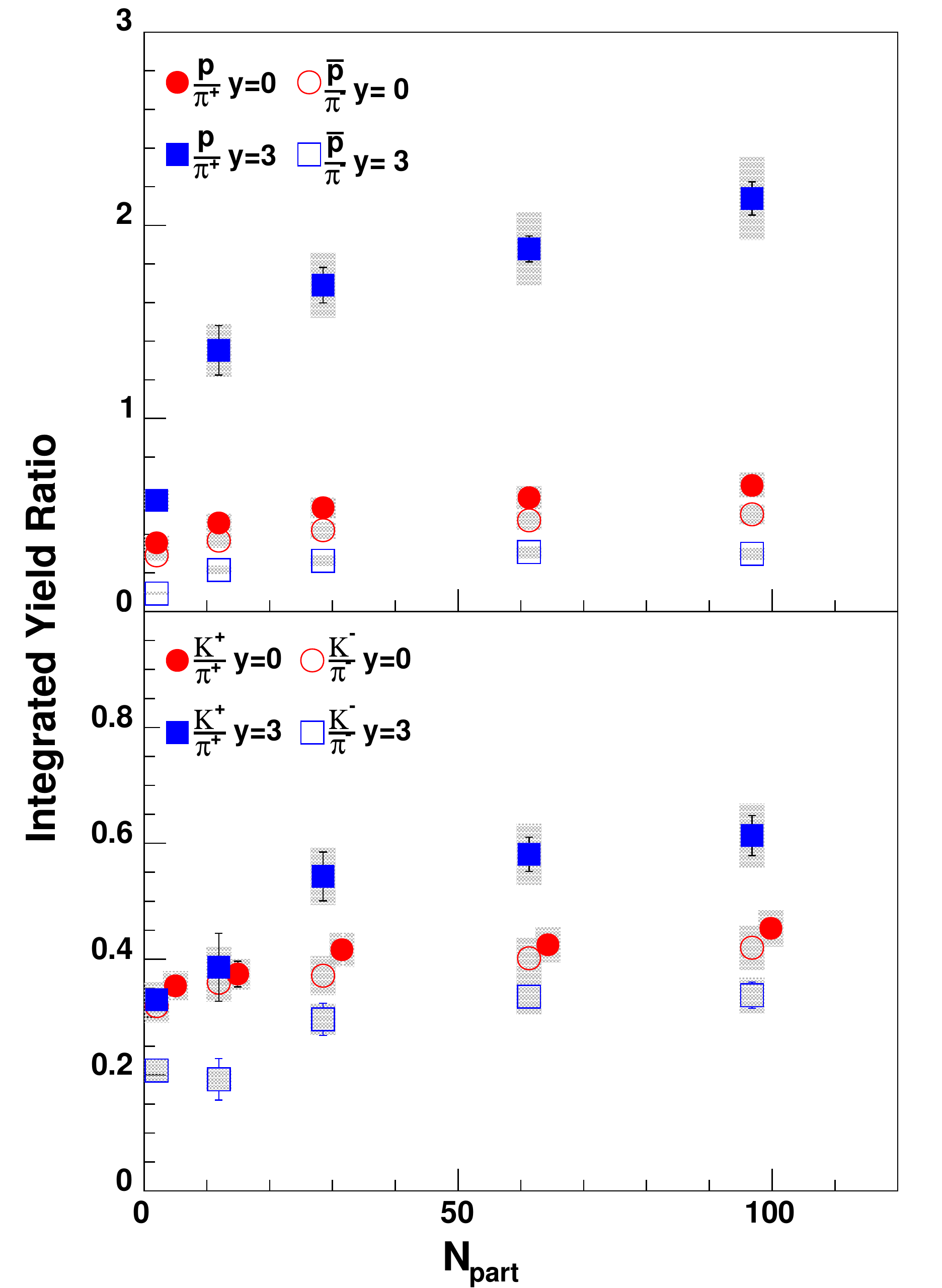}
\caption{(Color online) Ratio of particle yields,  $\frac{p}{\pi}$ (top) and $\frac{K}{\pi}$ (bottom) as a function of N$_{\rm part}$  for Cu+Cu and  and pp  collisions at $\sqrt{s_{NN}} = 200$ GeV. 
Before division the individual spectra have been 
integrated over $1.3 < p_T < 2.0$~GeV/$c$ for $y=0$    and over $1.5 < p_T < 3.0$~GeV/$c$ for $y=3$. Red circles denote  $y=0$   and blue squares $y=3$. Solid symbols represent ratios of positive particles while open symbols show the ratios of negative particles. The statistical errors are shown by bars while the systematic errors are shown by gray boxes. Note that the $\frac{K^+}{\pi^+}$ ratios at $y=0$ are displaced slightly for clarity. 
\label{ParticleRatioVsCent}}
\end{figure}

Both the $\frac{p}{\pi^{+}}$ and $\frac{\bar{p}}{\pi^{-}}$ ratios increase with $p_T$ at both rapidities
with  saturation taking place at $p_T \sim$ 1.6 GeV/$c$ for $y=3$.
The baryon-meson
ratios also show a moderate centrality dependence at the two
rapidities.
Both ratios at this \meanpt-range exceed the  maximum value of 0.2 observed in 
elementary $e^+e^-$ collisions at $\sqrt{s}$ = 91.2\,GeV for both quark and gluon jets  \cite{Abreu:1998vq, Abreu:2000nw}. 
Various mechanisms  such as quark coalescence, radial flow or baryon transport
dynamics may boost  the baryon-meson ratios at intermediate $p_T$ above the expected fragmentation value \cite{Guo:2000nz,Hirano:2003yp,Hirano:2003pw,Heinz:2001xi,Hwa:2004ng,Greco:2003mm,Greco:2004yc,Fries:2003kq,PhysRevC.65.041902,Vitev:2002wh,ToporPop:2002gf}. 
At mid-rapidity, the $p_T$ dependence of the BRAHMS $\frac{p}{\pi^+}$ ratio 
in central Au+Au collisions at $\sqrt{s_{NN}}$ = 200\,GeV \cite{Staszel:2005aw} 
has been reasonably described by recombination \cite{Hwa:2004ng}. 
Hydrodynamic models also qualitatively reproduced  the trend \cite{Hirano:2003yp,Hirano:2003pw,Heinz:2001xi}.
Although it is clear that the system size and the chemical properties of  the medium are important parameters, the detailed behavior of hadron 
production in the forward rapidity region remains a challenge to  microscopic models, as also seen in ref. 
\cite{Arsene:2009jg}.

The N$_{\rm part}$  dependence of the $\frac{K}{\pi}$ and $\frac{p}{\pi}$ ratios 
is displayed in  Fig.~\ref{ParticleRatioVsCent}. 
Here the individual spectra have been  integrated over the $p_T$ range  $1.3-2.0$~GeV/$c$ for $y=0$   and over $1.5-3.0$~GeV/$c$ for $y=3$. 
 We also show the ratios obtained from the BRAHMS $p+p$ data \cite{PhysRevLett.98.252001, pp200}. 
 The  $p+p$ values fit smoothly
with the trend of the lower $N_{part}$ values for Cu+Cu. 
The $\frac{K}{\pi}$ ratios increase slightly with N$_{\rm part}$, with the strongest centrality dependence seen for peripheral collisions at $y=3$.  The $\frac{K^{+}}{\pi^{+}}$ ratios are significantly larger than the 
$\frac{K^{-}}{\pi^{-}}$ ratios at $y=3$, but the two ratios are similar at $y=0$.
This   may be attributed to the larger baryon density at $y=3$ which increases the probability of associated  production for $K^{+}$.

The top panel of Fig.~\ref{ParticleRatioVsCent} shows the  integrated $\frac{p}{\pi^+}$ and  $\frac{\bar p}{\pi^-}$ ratios versus N$_{\rm part}$. 
The ratios seem to exhibit a monotonic increase with N$_{\rm part}$  
 at both rapidities. Again the pp ratios fit the trend of the lower N$_{\rm part}$ results but this dependence is stronger at $y=3$.
At mid-rapidity the ratios are 
smaller than unity with $\frac{p}{\pi^{+}}$ values slightly larger than the corresponding
 $\frac{\bar{p}}{\pi^{-}}$ values. 
At forward rapidity, the $\frac{p}{\pi^{+}}$ ratio is generally greater than unity and
is larger than the corresponding $\frac{\bar{p}}{\pi^{-}}$ ratio by  almost a factor of 6. 
This has also been observed in Au+Au collisions at the same center of mass energy per nucleon
\cite{Arsene:2009nr}. While the beam protons may be 
contributing to the $\frac{p}{\pi^{+}}$  ratio, the reason for such large differences  between the 
positive and negative baryon to meson ratios is not yet well understood.  
The increase of the baryon to meson ratios with centrality 
is consistent with trends exhibited by the $R_{AA}$ values, where
mesons become more suppressed for central events while baryons show only 
a weak if any centrality dependence.

\section{Summary and Conclusions}

The $\pi^{\pm}, K^{\pm},p,$ and $\bar{p}$ spectra from
Cu+Cu collisions at  $\sqrt{s_{NN}} = 200$ GeV  are well described by 
blast wave fits at both central and forward rapidities.  
As 
N$_{\rm part}$  
increases the kinetic temperature 
T$_{\rm kin}$ drops   
and the mean velocity $\langle \beta \rangle$ rises. For a given  $\langle \beta \rangle$, T$_{\rm kin}$ is about 15-20 MeV smaller at $y=3$ than at $y=0$. 
The particle yields per participant pair increase with N$_{\rm part}$. For a given 
N$_{\rm part}$ the kaon $dN/dy$ values are slightly larger in  
 Cu+Cu collisions than in Au+Au collisions.

Both pions and kaons from Cu+Cu collisions are suppressed relative to scaled $p+p$ collisions.  
The suppression is strongest for  central collisions as expected from models of parton energy loss or jet quenching. 
The suppression is slightly stronger at forward rapidity than at central rapidity 
suggesting  that the effect of the hot and dense medium extends to at least $y\approx 3$ at RHIC energies. This is despite the fact that 
 the rapidity densities in the forward region are about half of those at mid rapidity. 
The PHENIX collaboration has observed that increasing   
 parton energy loss with increasing beam energy can be compensated by hardening of the $p_T$ spectra, in such a way that $R_{AA}$ remains unchanged \cite{PhysRevLett.109.152301}.
A similar effect may be present when going to forward rapidities, so that the approximately
constant $R_{AA}$ can be a result of reduced energy loss combined with
steeper $p_T$ spectra for mesons. It is also possible that initial state effects such as nuclear shadowing are effecting particle production at forward rapidities,  \cite{Arsene:2004ux,Arsene:2006ts}. 

In contrast to the pions and kaons, protons with $p_T > 1.3 $ GeV/$c$  are enhanced relative to scaled $p+p$ collisions. 
 The baryon enhancement seen in $R_{AA}$   depends strongly on $p_T$ and rapidity but only weakly on centrality  
 The enhancement is similar for protons and antiprotons at  $y=0$, 
but is  stronger for antiprotons at forward rapidity. 
This is mainly because the $p+p$ reference spectrum for antiprotons at $y=3$ is much steeper than the corresponding proton spectrum \cite{PhysRevLett.98.252001}.

The 
 $\frac{\pi^{-}}{\pi^{+}}, \frac{K^-}{K^+}$ and $\frac{\bar p}{p}$  ratios  
 are almost independent of $p_T$ and centrality   
  but they do depend upon rapidity, presumably because of the higher net-baryon density in the forward region. 
 The $\frac{K^{\pm}}{\pi^{\pm}}, \frac{p}{\pi^+}$ and $\frac{\bar p}{\pi^-}$ ratios increase with N$_{\rm part}$ for 
 \pt\, up to $\simeq$ 1.6-2\,GeV/$c$  at both rapidities.
The four ratios at $y=3$ are seen to saturate for $p_T \ge 1.6$\,GeV/$c$.
At $y=3$, the kaon-pion and proton-pion ratios exhibit a
slightly different centrality dependence in the lowest
$N_{part}$ region.

At both rapidities the 
 $\frac{p}{\pi^{+}}$ and $\frac{\bar{p}}{\pi^{-}}$ ratios 
 in the intermediate $p_T$ region, i.e. 2.0\,GeV/$c$ $ < p_T < 3.5$\,GeV/$c$  
 are rather large for central collisions. 
This may be explained by either  
quark coalescence  
\cite{Guo:2000nz,Greco:2003mm,Fries:2003kq,Hwa:2004ng,Greco:2004yc}, radial flow\cite{Hirano:2003yp,Hirano:2003pw,Heinz:2001xi},
or baryon transport dynamics based on topological gluon field 
configurations
\cite{PhysRevC.65.041902,Vitev:2002wh,ToporPop:2002gf}. 
A similar baryon enhancement has been observed for Pb+Pb collisions at
$\sqrt{s_{\rm NN}}$ = 2.76 TeV \cite{Abelev:2014laa}. 
These data are also consistent with recombination  
\cite{Fries:2003kq} 
and hydrodynamical models \cite{Bozek:2012qs}.

Understanding the underlying mechanisms responsible for hadron production over the 
broad  range of 
transverse momentum and rapidity accessible at RHIC and providing a consistent description of all 
the various aspects of the hadron spectra in heavy ion collisions remains a major challenge. 
The current data will help  constrain theoretical attempts to reach such a synthesis.

\section{Acknowledgements}

This work was supported by the office of Nuclear Physics of the U.S. Department of energy,
the  Discovery Center of the Danish Natural Science Research Council, the Research Council of Norway, the Polish 
State Committee for Scientific Research (KBN), and the Romanian Ministry of Research.

\begin{table*}[ht] 
\begin{tabular}{|c|c|c|c|c|c|c|c|c|c|} \hline
\multicolumn{2}{|c|}{} & {\bf Cent.} & $\mathbf{\frac{dN}{dy}} $ & 
$\mathbf{{(\frac{dN}{dy})}_m}$ & $\mathbf{\frac{N_m}{N}} $ & $\mathbf{\langle p_{T}\rangle}$ {\bf (MeV/$c$)} & $\mathbf{\frac{\chi^2}{d.o.f}}$ & $\mathbf{n_0}$& {\bf T (MeV)}  \\ \hline
\multirow{8}{*}{$\pi^{+}$} & \multirow{4}{*}{y=0} &
$ 0-10\% $ &  81.1 $\pm$  3.1 $\pm$ 5.9  &  42.1 & 0.52 &    454 $\pm$     2 $\pm$ 21  &    0.2/  9 &  12.8 & 172\\ \cline{3-10}
 & & $ 10-30\% $ &  48.0 $\pm$  2.1 $\pm$ 3.5  &  24.3 & 0.51 &    445 $\pm$     4 $\pm$ 21  &    0.9/  9 &  11.6 & 164\\ \cline{3-10}
 & & $ 30-50\% $ &  21.8 $\pm$  0.5 $\pm$ 1.6  &  10.8 & 0.50 &    438 $\pm$     2 $\pm$ 21  &    0.2/  9 &  11.2 & 159\\ \cline{3-10}
 & & $ 50-70\% $ &   8.5 $\pm$  0.40 $\pm$ 0.62  &   4.0 & 0.47 &    418 $\pm$     5 $\pm$ 20  &    2.4/  9 &  10.2 & 147\\ \cline{2-10}
& \multirow{4}{*}{y=3} &
$ 0-10\% $ &  33.7 $\pm$  3.5 $\pm$ 2.4  &  11.2 & 0.33 &    401 $\pm$     8 $\pm$ 19  &   12.2/  9 &  17.0 & 159\\ \cline{3-10}
 & & $ 10-30\% $ &  18.3 $\pm$  1.9 $\pm$ 1.3  &   6.6 & 0.36 &    424 $\pm$     9 $\pm$ 20  &   16.2/  9 &  19.3 & 173\\ \cline{3-10}
 & & $ 30-50\% $ &   9.3 $\pm$  0.88 $\pm$ 0.67  &   3.1 & 0.33 &    403 $\pm$     7 $\pm$ 19  &    7.3/  9 &  16.1 & 158\\ \cline{3-10}
 & & $ 50-70\% $ &   3.3 $\pm$  0.70 $\pm$ 0.24  &   1.2 & 0.35 &    418 $\pm$    14 $\pm$ 20  &   23.0/  9 &  17.9 & 168\\ \cline{2-10}
 \hline \hline
\multirow{8}{*}{$\pi^{-}$} & \multirow{4}{*}{y=0} &
$ 0-10\% $ &  78.0 $\pm$  3.3 $\pm$ 4.9  &  41.1 & 0.53 &    460 $\pm$     4 $\pm$ 22  &    0.9/  9 &  13.3 & 176\\ \cline{3-10}
 & & $ 10-30\% $ &  44.7 $\pm$  1.9 $\pm$ 2.8  &  23.2 & 0.52 &    455 $\pm$     5 $\pm$ 21  &    2.1/  9 &  12.3 & 170\\ \cline{3-10}
 & & $ 30-50\% $ &  20.5 $\pm$  0.9 $\pm$ 1.3  &  10.2 & 0.50 &    441 $\pm$     3 $\pm$ 21  &    0.5/  9 &  10.6 & 158\\ \cline{3-10}
 & & $ 50-70\% $ &   8.0 $\pm$  0.36 $\pm$ 0.51  &   3.8 & 0.47 &    421 $\pm$     4 $\pm$ 20  &    0.7/  9 &  10.2 & 148\\ \cline{2-10}
& \multirow{4}{*}{y=3} &
$ 0-10\% $ &  32.4 $\pm$  3.1 $\pm$ 2.3  &  11.2 & 0.35 &    411 $\pm$     8 $\pm$ 19  &   14.5/  9 &  17.5 & 164\\ \cline{3-10}
 & & $ 10-30\% $ &  20.8 $\pm$  1.8 $\pm$ 1.5  &   7.4 & 0.36 &    419 $\pm$     8 $\pm$ 20  &   13.9/  9 &  21.0 & 173\\ \cline{3-10}
 & & $ 30-50\% $ &  11.1 $\pm$  1.4 $\pm$ 0.8  &   3.5 & 0.32 &    392 $\pm$     9 $\pm$ 18  &   16.8/  9 &  15.5 & 152\\ \cline{3-10}
 & & $ 50-70\% $ &   3.6 $\pm$  0.40 $\pm$ 0.26  &   1.3 & 0.36 &    424 $\pm$     8 $\pm$ 20  &    5.2/  9 &  20.5 & 174\\ \cline{2-10}
 \hline
\end{tabular}

\caption{Extracted fit results for pions based on a Levy function. 
The systematic uncertainty estimate follows the statistical error.}
\label{table:MrsPions}
\end{table*}


\begin{table*}[ht] 
\begin{tabular}{|c|c|r|c|  l|  l|  l|  l|  c|} \hline 
\multicolumn{2}{|c|}{} & {\bf Cent.} & ${\frac{dN}{dy}}$  & 
$\mathbf{{(\frac{dN}{dy})}_m}$ & $\mathbf{\frac{N_m}{N}} $ & $\mathbf{\langle p_{T}\rangle}$ {\bf (MeV/$c$)} & $\mathbf{\frac{\chi^2}{d.o.f}}$ & {\bf T (MeV)}  \\ \hline
\multirow{8}{*}{$K^{+}$} & \multirow{4}{*}{y=0} &
$ 0-10\% $ &  12.3 $\pm$  0.32 $\pm$ 0.89  &   ~7.6 & 0.62 &    674 $\pm$    10 $\pm$ 22  &    1.6/7 & 277\\ \cline{3-9}
 & & $ 10-30\% $ &   ~6.9 $\pm$  0.01 $\pm$ ~0.50  &   ~4.2 & 0.61 &    663 $\pm$     7 $\pm$ 21  &    0.9/7 & 271\\ \cline{3-9}
 & & $ 30-50\% $ &   ~2.8 $\pm$  0.02 $\pm$ ~0.20  &   ~1.7 & 0.62 &    667 $\pm$    14 $\pm$ 21  &    3.9/7 & 273\\ \cline{3-9}
 & & $ 50-70\% $ &   ~1.0 $\pm$  0.05 $\pm$ ~0.12  &   ~0.6 & 0.59 &    625 $\pm$    14 $\pm$ 20  &    3.4/7 & 251\\ \cline{2-9}
& \multirow{4}{*}{y=3} &
$ 0-10\% $           &   ~4.6 $\pm$  0.29 $\pm$ 0.33  &   ~1.3 & 0.27 &    611 $\pm$    14 $\pm$ 20  &    4.1/4 & 244\\ \cline{3-9}
 & & $ 10-30\% $ &   ~3.0 $\pm$  0.20 $\pm$ ~0.22  &   ~0.78 & 0.26 &    594 $\pm$    19 $\pm$ 19  &    5.8/4 & 235\\ \cline{3-9}
 & & $ 30-50\% $ &   ~1.4 $\pm$  0.11 $\pm$ ~0.10  &   ~0.34 & 0.25 &    577 $\pm$    20 $\pm$ 18  &    5.7/4 & 226\\ \cline{3-9}
 & & $ 50-70\% $ &   ~0.39 $\pm$  0.05 $\pm$ ~0.03  &   ~0.10 & 0.26 &    600 $\pm$    27 $\pm$ 19  &    5.4/4 & 238\\ \cline{2-9}
 \hline \hline
\multirow{8}{*}{$K^{-}$} & \multirow{4}{*}{y=0} &
$ 0-10\% $           &  11.2 $\pm$  0.23 $\pm$ 0.71  &   ~7.2 & 0.64 &    682 $\pm$     9 $\pm$ 22  &    2.0/ 8 & 282\\ \cline{3-9}
 & & $ 10-30\% $ &   ~6.1 $\pm$  0.15 $\pm$ 0.38  &   ~3.9 & 0.64 &    683 $\pm$    12 $\pm$ 22  &    4.3/8 & 282\\ \cline{3-9}
 & & $ 30-50\% $ &   ~2.5 $\pm$  0.08 $\pm$ 0.16  &   ~1.6 & 0.63 &    677 $\pm$    26 $\pm$ 22  &   11.4/8 & 279\\ \cline{3-9}
 & & $ 50-70\% $ &   ~0.7 $\pm$  0.02 $\pm$ 0.10  &   ~0.5 & 0.64 &    685 $\pm$    28 $\pm$ 22  &   10.8/8 & 283\\ \cline{2-9}
& \multirow{4}{*}{y=3} &
$ 0-10\% $ &   ~3.9 $\pm$  0.02 $\pm$ 0.28           &   ~0.96 & 0.25 &    569 $\pm$    12 $\pm$ 18  &    5.4/5 & 222\\ \cline{3-9}
 & & $ 10-30\% $ &   ~2.2 $\pm$  0.12 $\pm$ 0.16  &   ~0.57 & 0.26 &    580 $\pm$    10 $\pm$ 19  &    4.4/5 & 227\\ \cline{3-9}
 & & $ 30-50\% $ &   ~1.0 $\pm$  0.05 $\pm$ 0.07  &   ~0.24 & 0.23 &    551 $\pm$    12 $\pm$ 18  &    4.8/5 & 213\\ \cline{3-9}
 & & $ 50-70\% $ &   ~0.34 $\pm$  0.02 $\pm$ 0.03  &   ~0.08 & 0.25 &    572 $\pm$    12 $\pm$ 18  &    1.5/5 & 223\\ \cline{2-9}
 \hline
\end{tabular}

\caption{Extracted fit results for kaons based on an exponential function in $m_{T}$.  The systematic uncertainty estimate follows the statistical error.}
\label{table:MrsKaons}
\end{table*}

\begin{table*}[ht] 
\begin{tabular}{|c|c|c|c|c|c|c|c|c|} \hline
\multicolumn{2}{|c|}{} & {\bf Cent.} & $\mathbf{\frac{dN}{dy}} $ & 
$\mathbf{{(\frac{dN}{dy})}_m}$ & $\mathbf{\frac{N_m}{N}} $ & $\mathbf{\langle p_{T}\rangle}$ {\bf (MeV/$c$)} & $\mathbf{\frac{\chi^2}{d.o.f}}$ & {\bf T (MeV)}  \\ \hline
\multirow{8}{*}{$p$} & \multirow{4}{*}{y=0} &
$ 0-10\% $ &   8.1 $\pm$  0.03 $\pm$ 0.51  &   5.7 & 0.70 &    896 $\pm$    18 $\pm$ 29  &    9.1/  9 & 332\\ \cline{3-9}
 & & $ 10-30\% $ &   4.7 $\pm$  0.10 $\pm$ 0.29  &   3.2 & 0.69 &    874 $\pm$     9 $\pm$ 28  &    1.9/  9 & 320\\ \cline{3-9}
 & & $ 30-50\% $ &   2.1 $\pm$  0.05 $\pm$ 0.13  &   1.4 & 0.67 &    831 $\pm$    14 $\pm$ 27  &    6.4/  9 & 296\\ \cline{3-9}
 & & $ 50-70\% $ &   0.7 $\pm$  0.03 $\pm$ 0.05  &   0.46 & 0.64 &    784 $\pm$    25 $\pm$ 25  &   12.3/  9 & 271\\ \cline{2-9}
& \multirow{4}{*}{y=3} &
$ 0-10\% $ &   7.0 $\pm$  0.03 $\pm$ 0.44           &   5.1 & 0.74 &    775 $\pm$    13 $\pm$ 25  &   10.3/ 10 & 266\\ \cline{3-9}
 & & $ 10-30\% $ &   4.3 $\pm$  0.11 $\pm$ 0.27  &   3.2 & 0.73 &    761 $\pm$    16 $\pm$ 24  &   12.4/ 10 & 259\\ \cline{3-9}
 & & $ 30-50\% $ &   2.0 $\pm$  0.07 $\pm$ 0.12  &   1.4 & 0.71 &    737 $\pm$    38 $\pm$ 24  &   23.2/ 10 & 247\\ \cline{3-9}
 & & $ 50-70\% $ &   0.76 $\pm$  0.04 $\pm$ 0.05  &   0.54 & 0.70 &    712 $\pm$    57 $\pm$ 23  &   30.7/ 10 & 234\\ \cline{2-9}
 \hline \hline
\multirow{8}{*}{$\bar{p}$} & \multirow{4}{*}{y=0} &
$ 0-10\% $ &   6.0 $\pm$  0.17 $\pm$ 0.38            &   4.3 & 0.70 &    906 $\pm$    38 $\pm$ 29  &   15.9/  9 & 338\\ \cline{3-9}
 & & $ 10-30\% $ &   3.5 $\pm$  0.11 $\pm$ 0.22  &   2.4 & 0.69 &    880 $\pm$    12 $\pm$ 28  &    3.9/  9 & 323\\ \cline{3-9}
 & & $ 30-50\% $ &   1.5 $\pm$  0.04 $\pm$ 0.10  &   1.0 & 0.68 &    839 $\pm$    16 $\pm$ 27  &    7.7/  9 & 300\\ \cline{3-9}
 & & $ 50-70\% $ &   0.6 $\pm$  0.02 $\pm$ 0.04  &   0.38 & 0.64 &    781 $\pm$    22 $\pm$ 25  &   10.9/  9 & 269\\ \cline{2-9}
& \multirow{4}{*}{y=3} &
$ 0-10\% $ &   1.2 $\pm$  0.04 $\pm$ 0.07            &   0.73 & 0.62 &    750 $\pm$    20 $\pm$ 24  &   11.7/  9 & 254\\ \cline{3-9}
 & & $ 10-30\% $ &   0.86 $\pm$  0.04 $\pm$ 0.05  &   0.53 & 0.61 &    731 $\pm$    31 $\pm$ 23  &   17.1/  9 & 244\\ \cline{3-9}
 & & $ 30-50\% $ &   0.37 $\pm$  0.01 $\pm$ 0.02  &   0.22 & 0.60 &    719 $\pm$    27 $\pm$ 23  &   14.2/  9 & 238\\ \cline{3-9}
 & & $ 50-70\% $ &   0.15 $\pm$  0.01 $\pm$ 0.01  &   0.09 & 0.57 &    685 $\pm$    42 $\pm$ 22  &   18.2/  9 & 221\\ \cline{2-9}
 \hline
\end{tabular}

\caption{Extracted fit results for protons and anti-protons based on an exponential function in $m_{T}$.  The systematic uncertainty estimate follows the statistical error.}
\label{table:MrsProtons}
\end{table*}

\clearpage

\bibliography{BRAHMS_CuCuRefs-1}
\end{document}